\documentclass[11pt]{article}

\usepackage{amsmath}
\usepackage{amssymb}

\usepackage{cancel}
\usepackage{graphicx}
\usepackage{hyperref}
\usepackage{subfigure}

\newcommand{\R}{{\mathbb{R}}}

\newcommand{\eps}{\varepsilon}
\renewcommand{\vec}{\mathbf}
\newcommand{\BigO}{\mathcal O}

\makeatletter
\def\captionof#1#2{{\def\@captype{#1}#2}}
\makeatother

\newtheorem{theorem}{Theorem}[section]
\newtheorem{lmm}[theorem]{Lemma}
\newtheorem{rmrk}[theorem]{Remark}

\numberwithin{equation}{section}

\begin{document}

\title{A Mathematical Description of the IDSA for Supernova Neutrino transport, its discretization and a comparison with a finite volume scheme for Boltzmann's Equation}

\author{
{H.~Berninger}\thanks{Universit\'e de Gen\`eve, Section de Math\'ematiques, 2-4,~rue du Li\`evre, CP~64, CH-1211~Gen\`eve}
\and {E.~Fr\'enod}\thanks{Universit\'e de Bretagne-Sud, Laboratoire de Math\'ematiques, Centre Yves Coppens, Bat.~B, BP~573, F-56017~Vannes}
\and {M.J.~Gander}\thanks{Universit\'e de Gen\`eve, Section de Math\'ematiques, 2-4,~rue du Li\`evre, CP~64, CH-1211~Gen\`eve}
\and {M.~Liebend\"orfer}\thanks{Universit\"at Basel, Departement Physik, Klingelbergstrasse~82, CH-4056~Basel}
\and {J.~Michaud}\thanks{Universit\'e de Gen\`eve, Section de Math\'ematiques, 2-4,~rue du Li\`evre, CP~64, CH-1211~Gen\`eve}
\and {N.~Vasset}\thanks{Universit\"at Basel, Departement Physik, Klingelbergstrasse~82, CH-4056~Basel}
}

\maketitle

{\bf Abstract - }{\small
In this paper we give an introduction to the Boltzmann equation for neutrino transport used in core collapse supernova models as well as a detailed mathematical description of the \emph{Isotropic Diffusion Source Approximation} (IDSA) established in \cite{LiebendoerferEtAl09}. Furthermore, we present a numerical treatment of a reduced Boltzmann model problem based on time splitting and finite volumes and revise the discretization of the IDSA in \cite{LiebendoerferEtAl09} for this problem. Discretization error studies carried out on the reduced Boltzmann model problem and on the IDSA show that the errors are of order one in both cases. By a numerical example, a detailed comparison of the reduced model and the IDSA is carried out and interpreted. For this example the IDSA modeling error with respect to the reduced Boltzmann model is numerically determined and localized.}
\\ ~\\ 
{\bf R\'esum\'e - } {\small
Dans cet article, nous donnons une introduction \`a l'\'equation de Boltzmann pour le transport des neutrinos dans les mod\`eles de supernovae \`a effondrement de coeur ainsi qu'une description d\'etaill\'ee de l'\emph{Isotropic Diffusion Source Approximation} (IDSA) d\'evelopp\'ee dans \cite{LiebendoerferEtAl09}. De plus, nous pr\'esentons le traitement num\'erique d'un mod\`ele de Boltzmann simplifi\'e bas\'e sur une d\'ecomposition en temps de l'op\'erateur et sur un algorithme de volumes finis ainsi que l'adaptation de la discr\'etisation de l'IDSA de \cite{LiebendoerferEtAl09} \`a notre mod\`ele. Les \'etudes de l'erreur de discr\'etisation faites sur le mod\`ele de Boltzmann simplifi\'e et sur l'IDSA montrent que les erreurs sont d'ordre un dans les deux cas. A l'aide d'un exemple num\'erique, nous comparons et interpr\'etons en d\'etail les deux mod\`eles. Pour cet exemple, l'erreur de mod\'elisation de l'IDSA par rapport au mod\`ele de Boltzmann simplifi\'e est d\'etermin\'ee num\'eriquement et localis\'ee.}

\section*{Introduction}

Modelling neutrino transport is crucial for the simulation of core collapse supernovae since more than $99\%$ of the released gravitational binding energy is carried away by neutrinos \cite[p.\;361]{RamppJanka02} that are also assumed to feed the shock leading to the explosion. However, full 3D Boltzmann neutrino transport models are still computationally too costly to solve. The Isotropic Diffusion Source Approximation (IDSA) intends to capture the most important processes of neutrino transport while being computationally feasible \cite{LiebendoerferEtAl09}. The main idea of the IDSA is to consider
an additive decomposition of the neutrino distribution function into a trapped and a free streaming particle component. The resulting equations related to these two components are supposed to be coupled by a source term. Both the source term and the equations which are reduced to the main physical properties of the two particle components are derived. The source term is based on a diffusion limit and, for non-diffusive regimes, limited from above and below on the basis of the free streaming and reaction limits.

In the first part of this paper, we give an introduction to the $\BigO(v/c)$ Boltzmann equation for radiative transfer in the comoving frame (Section~\ref{radiative}) as well as to the IDSA of this model (Section~\ref{IDSA}). This part is based on the presentation in \cite{LiebendoerferEtAl09}, however, it is more mathematically oriented and, in particular, the derivation of the diffusion source, which is only sketched in \cite{LiebendoerferEtAl09}, is presented in a comprehensive manner. As in \cite{LiebendoerferEtAl09}, we restrict ourselves to the spherically symmetric case here, however, both models can be extended to the nonsymmetric 3D case. 

In the second part of the paper, we introduce a reduced Boltzmann model problem in which we mainly assume frozen background matter and give the corresponding IDSA (Section~\ref{reducedSec}). For this reduced model we present a numerical solution technique based on time splitting between the reaction and the transport part. While the reaction part has an analytical solution, a conservative formulation can be found for the transport part for which a finite volume scheme is established (Section~\ref{numericaltreatSec}). Here we also recall the discretization of the IDSA given in \cite{LiebendoerferEtAl09} and adapt it to the reduced model. On this basis, we present several numerical results (Section~\ref{numericalresSec}). To start with, convergence studies are carried out on the reduced Boltzmann model problem and on the IDSA in order to have realistic estimates of the discretization error. The results show that the errors are of order one in both cases. Then we establish a numerical example in order to compare the behaviour of solutions of the reduced model and of the IDSA. Extensive interpretations of the solutions in reaction and free streaming regimes and in the transition regime are given. For this example we also find a considerable IDSA modeling error with respect to the reduced Boltzmann model numerically and show that the error is mainly localized in the transition regime.

\section{Boltzmann's radiative transfer equation}
\label{radiative}

The $\BigO(v/c)$ Boltzmann's equation is a widely accepted model for the radiative transfer of neutrinos in core collapse supernovae, see, e.g., \cite{Bruenn85,MezzacappaBruenn93a,RamppJanka02,LiebendoerferEtAl09}. 
We present the version used in \cite{LiebendoerferEtAl09} and write it in short as
\begin{equation}
\underbrace{\frac{1}{c}\frac{d f}{d t}+ \mu \frac{\partial f}{\partial r} +F_\mu{}\frac{\partial f}{\partial \mu}+F_\omega{}\frac{\partial f}{\partial \omega}}_{\mathcal D(f)} = j{} \underbrace{-\tilde\chi{} f+  \mathcal C(f)}_{\mathcal J(f)}\,. \label{Boltz}
\end{equation}

Equation \eqref{Boltz} represents the special relativistic transport equation for massless fermions up to the order $\BigO(v/c)$, i.e., the neutrinos are considered to move with light speed $c$ while the background matter moves with velocity $v$. We refer to \cite[\S 95]{Mihalas84} for a derivation of \eqref{Boltz} and \cite{RamppJanka02} for a discussion of possibly additional $\BigO(v/c)$ terms that might have to be considered.

The equation describes the evolution of the distribution function 
\begin{equation}
\label{fDef}
f:[0,T]\times (0,R]\times [-1,1]\times (0,E]\to [0,1]\,,\quad (t,r,\mu,\omega)\mapsto f(t,r,\mu,\omega)\,,
\end{equation}
of the neutrinos. Here, $T>0$ is some end time, $R>0$ some maximal radius and $E>0$ some maximal neutrino energy. We consider \eqref{Boltz} to be given in spherical symmetry, i.e., $(0,R]$ represents the spatial domain $\Omega\subset\R^3$, which is the ball around the origin with radius~$R$. The variable $\mu\in [-1,1]$ is the cosine of the angle between the outward radial direction and the direction of neutrino propagation and $\omega\in\R^+$ is the neutrino energy, both (the phase space variables) given in the frame comoving with the background matter, cf.~\cite[p.~640]{MezzacappaBruenn93a}. 

With regard to the notation we use the Lagrangian time derivative 
$$
\frac{d}{d t} = \frac{\partial}{\partial t} + v\frac{\partial}{\partial r}\,,
$$ 
in the comoving frame. Furthermore, we have 
$$
F_\mu{} = F_\mu^0 + F_\mu^1{}\,,\quad F_\mu^0 =\frac{1}{r}(1-\mu^2)\,,\quad F_\mu^1{}=\mu \left( \frac{d \ln\rho}{cd t}+\frac{3v}{cr}\right)(1-\mu^2)\,,
$$ 
where $F_\mu^0\frac{\partial f}{\partial \mu}$ accounts for the change in propagation direction due to inward or outward movement of the neutrino. The second term $F_\mu^1{}\frac{\partial f}{\partial \mu}$ represents the aberration (i.e., the change observed in the comoving frame) of the neutrino propagation due to the motion of the background matter with the density $\rho$. Finally, the product of the factor 
$$
F_\omega{} =\left[\mu^2 \left( \frac{d \ln\rho}{cd t}+\frac{3v}{cr}\right)-\frac{v}{cr} \right]\omega\, ,
$$ 
with $\frac{\partial f}{\partial \omega}$ accounts for the (Doppler--)shift in neutrino energy due to the motion of the matter. The left hand side of the Boltzmann equation is abbreviated by $\mathcal{D}(f)$ where $\mathcal{D}$ is a linear operator.

The dependency of $\mathcal{D}$ on the background matter occurring in the comoving frame vanishes in case of a static background with frozen matter where one can pass to the laboratory frame by setting $F_\mu^1{}=F_\omega{}=0$, i.e., 
\begin{equation}
\label{Boltz_frozen}
\frac{1}{c}\frac{\partial f}{\partial t}+ \mu \frac{\partial f}{\partial r} +\frac{1}{r}(1-\mu^2)\frac{\partial f}{\partial \mu} 
= j{}-\tilde\chi{} f+  \mathcal C(f)\,,
\end{equation}
and using Lorentz transformed quantities in \eqref{Boltz_frozen}, see \cite[\S 90,\;\S 95]{Mihalas84}.

Although in the infall phase \cite[p.~787]{Bruenn85} it is enough to consider only electron neutrinos $\nu_e$, for postbounce simulations \cite[p.~1179]{LiebendoerferEtAl09} one needs at least two Boltzmann equations in order to obtain the transport of both electron neutrinos $\nu_e$ and electron antineutrinos $\bar\nu_e$. In general, one needs to include muon and tau neutrinos and their antiparticles, too \cite{BurasEtAl03}. All different types of neutrinos are transported independently, so that in general one has to deal with up to six Boltzmann equations that are, however, coupled via their right hand sides. Since all these equations have the same basic structure, it is enough for our purpose to consider only one Boltzmann equation as a prototype.

The source and sink terms on the right hand side of \eqref{Boltz} account for interaction of neutrinos with background matter. They include emission and absorption
\begin{equation*}
\label{reactions}
\begin{aligned}
e^-+p&\rightleftharpoons n+\nu_e\, ,\\
e^++n&\rightleftharpoons p+\bar\nu_e\, ,
\end{aligned}
\end{equation*}
of electron neutrinos $\nu_e$ or electron antineutrinos $\bar\nu_e$ by protons $p$ or neutrons $n$, the forward reactions being known as electron $e^-$ or positron $e^+$ capture. Analogous reactions occur in case of electron $e^-$ or positron $e^+$ capture of nuclei. They depend on the state of the background matter and result in neutrino emissivity $j(\omega)$ and absorptivity $\chi(\omega)$ whose sum is the neutrino opacity $\tilde\chi=j+\chi$ in \eqref{Boltz}, cf.~\cite{MezzacappaBruenn93a} and \cite[p.~1177]{LiebendoerferEtAl09}. Concrete formulas for $j(\omega)$ and $\chi(\omega)$, which are nonlinear in $\omega$, have been derived in \cite[pp.~822--826]{Bruenn85} for both electron neutrino and its antiparticle.

By the term $\mathcal C(f)$, the right hand side of \eqref{Boltz} also accounts for isoenergetic scattering of neutrinos (or antineutrinos) on protons, neutrons and nuclei. $\mathcal C(f)$ is a linear integral operator in $f$, the so called collision integral, and is given by
\begin{equation}\label{isoscattering}
\mathcal C(f(t,r,\mu,\omega))
=\frac{\omega^2}{c(hc)^3}\left[\int_{-1}^{1} R(\mu,\mu',\omega) \left(f(t,r,\mu',\omega)-f(t,r,\mu,\omega)\right) d\mu'\right], 
\end{equation}
where the isoenergetic scattering kernel $R(\mu,\mu',\omega)$ is symmetric in $\mu$ and $\mu'$ and depends nonlinearly on all its entries, see \cite[pp.~806/7,\,826--828]{Bruenn85} for concrete formulas that also exhibit the dependency of $R$ on the background matter. In \eqref{isoscattering} Planck's constant is denoted by~$h$, the term corresponding to $f(t,r,\mu',\omega)$ accounts for in-scattering while the term corresponding to $f(t,r,\mu,\omega)$ represents out-scattering \cite[p.~698]{LiebendoerferEtAl09}. Note that if $f$ does not depend on $\mu$ we have $\mathcal{C}(f)=0$. As an immediate consequence of the symmetry of the collision kernel with respect to $\mu$ and $\mu'$ we obtain for any $f$
\begin{equation}
\label{collision_vanishes}
\int_{-1}^1\mathcal{C}(f)\,d\mu=0\,.
\end{equation}

Further possible source terms stemming from neutrino interactions with the background matter such as, e.g., neutrino electron scattering (cf.~\cite[p.~774]{Bruenn85}), are neglected in \cite{LiebendoerferEtAl09}. 
We abbreviate the right hand side of Boltzmann's equation \eqref{Boltz} by $j+\mathcal J(f)$ where $\mathcal J(f)$ is linear in $f$.

\section{Isotropic Diffusion Source Approximation (IDSA)}
\label{IDSA}

In this section we give an introduction to the Isotropic Diffusion Source Approximation (IDSA) that has been developed in \cite{LiebendoerferEtAl09}. The aim of this approximation of Boltzmann's equation \eqref{Boltz} is to reduce the computational cost for its solution, making use of the fact that \eqref{Boltz} is mainly governed by diffusion of neutrinos in the inner core and by transport of free streaming neutrinos in the outer layers of a star. The following ansatz for the IDSA intends to avoid the solution of the full Boltzmann equation in a third domain in between these two regimes as well as the detection of the corresponding domain boundaries.

\subsection{Ansatz: Decomposition into trapped and streaming neutrinos}

We assume a \emph{decomposition} of
\begin{equation*}
f= f^t + f^s
\end{equation*}
\emph{on the whole domain} into distribution functions $f^t$ and $f^s$ supposed to account for \emph{trapped} and for \emph{streaming} neutrinos, respectively.

With this assumption and by linearity $\mathcal D(f) = \mathcal D(f^t) + \mathcal D(f^s)$ and $\mathcal J(f) = \mathcal J(f^t)+ \mathcal J(f^s)$, solving the Boltzmann equation \eqref{Boltz}, i.e., 
\begin{equation}
\label{Boltz_pure}
\mathcal D(f)=j+\mathcal J(f)\,,
\end{equation}
is equivalent to solving the two equations
\begin{align}
\mathcal D(f^t) &= j+\mathcal J (f^t) -\Sigma \, ,\label{decomposedtrapped}\\[2mm]
\mathcal D(f^s) &= \mathcal J (f^s) +\Sigma \, ,\label{decomposedstreaming}
\end{align}
with an arbitrary coupling term $\Sigma=\Sigma(t,r,\mu,\omega,f^t,f^s,j,\mathcal{J})$. For the IDSA one establishes approximations of the angular mean of these two equations arising from physical properties of trapped and streaming particle, respectively, and one determines an appropriate coupling function $\Sigma(t,r,\mu,\omega,f^t,f^s,j,\mathcal{J})$.

\subsection{Hypotheses and their consequences for trapped and streaming particle equations}\label{Hyp}

Concretely, one uses the following hypotheses. First, the trapped particle component $f^t$ as well as $\Sigma$ are assumed to be \emph{isotropic}, i.e., independent of $\mu$. Taking the angular mean of equation  \eqref{decomposedtrapped} this immediately leads to the trapped particle equation
\begin{equation}
\label{eq:trapped}
\frac{d f^t}{cd t} + \frac{1}{3}\frac{d \ln\rho}{cd t}\omega\frac{\partial f^t}{\partial \omega} = j-\tilde \chi f^t -\Sigma \, ,
\end{equation}
in which we slightly abuse the notation 
\begin{equation}
\label{f_t_is_beta_t}
f^t=\frac{1}{2}\int_{-1}^1 f^t d\mu\, ,
\end{equation}
concerning the domain of definition of the isotropic $f^t$ given in~\eqref{fDef}. The isotropic source term $\Sigma$ is treated in the same way. Here, and in what follows, we always assume that we can interchange the integral and the differentiation operators.

Next, $f^t$ is assumed to be in the \emph{diffusion limit}, which is physically at least justified for the inner core of the star. 
In order to derive the diffusion limit, a Legendre expansion of the scattering kernel $R(\mu,\mu',\omega)$ with respect to its angular dependence, truncated after the second term, is used in \cite[App.~A]{LiebendoerferEtAl09} for an approximation of the collision integral, see Subsection~\ref{Legendre}. In fact, this approximation is essential for the derivation of the diffusion limit and thus the corresponding definition of $\Sigma$ that we will provide in Subsection~\ref{diffusion_limit}.

Second, the streaming particle component $f^s$ is assumed to be in the \emph{free streaming limit}. This justifies to neglect the collision integral in \eqref{decomposedstreaming}, which by \eqref{collision_vanishes}, however, vanishes anyway after angular integration. Furthermore, it justifies to neglect the dynamics of background matter so that one can use the laboratory frame formulation \eqref{Boltz_frozen} of \eqref{Boltz} with frozen matter (here, we also neglect the Lorentz transformation). For the same reason one can assume the free streaming particle to be in the \emph{stationary state limit} and drop the time derivative in~\eqref{Boltz_frozen} which then, after angular integration, becomes the streaming particle equation
\begin{equation}
\label{eq:streaming}
\frac{1}{r^2}\frac{\partial}{\partial r}\left( \frac{r^2}{2}\int_{-1}^1f^s\mu d\mu\right) = -\frac{\tilde\chi}{2}\int_{-1}^1f^s d\mu+\Sigma\,.
\end{equation}
Since in spherical symmetry, with the radial unit vector $\mathbf{e}_r$, the gradient and the divergence operators are given by 
\begin{equation}
\label{nabla_div_in_r}
\nabla\psi(r)=\frac{\partial }{\partial r}\psi(r)\,\mathbf{e}_r\qquad\mbox{and}\qquad\nabla\cdot \vec F(r)=\frac{1}{r^2}\frac{\partial }{\partial r}\left(r^2 F_r(r)\right)\,,\\[1mm]
\end{equation}
for scalar fields $\psi:\R\to\R$ and vector fields $\vec F=(F_r,0,0):\R\to\R^3$ (see, e.g., \cite[p.~679]{Mihalas84}), 
this equation can be reformulated as a Poisson equation 
for a spatial potential $\psi$ of the first angular moment of $f^s$.

For the solution of \eqref{eq:streaming}, the approximate relationship
\begin{equation}
\label{scattering_sphere}
\int_{-1}^1f^s(\omega) \mu d\mu
=\left(1+\sqrt{1-\left(\frac{R_\nu(\omega)}{\max(r,R_\nu(\omega))}\right)^2}\right)\frac{1}{2}\int_{-1}^1f^s(\omega) d\mu\, ,
\end{equation}
between the particle flux and the particle density of streaming neutrinos, which has been suggested by Bruenn in \cite{LiebendoerferEtAl04}, is applied. Here, $R_\nu(\omega)>0$ is the energy dependent radius of the neutrino scattering spheres. In addition to spherical symmetry, this approximation is based on the assumption that all free streaming particles of a given energy $\omega$ are emitted isotropically at their corresponding scattering sphere \cite[p.~1178]{LiebendoerferEtAl09}. As a consequence, the flux can be expressed as a product of the particle density and the geometrical factor in \eqref{scattering_sphere}. Note that this factor is equal to $1$ when $r \leq R_{\nu}(\omega)$, which follows from the isotropy of $f$ inside the scattering spheres, and increases up to $2$ in the limit $r \rightarrow \infty$. The latter expresses the fact that the neutrinos tend to stream radially outwards so that the distribution function $f$ accumulates at $\mu = 1$.

\subsection{Legendre expansion of the scattering kernel}
\label{Legendre}

As mentioned in the last subsection, we now seek for an approximation of the collision integral by a Legendre expansion of the scattering kernel. For an introduction to Legendre expansions by spherical harmonics we refer to \cite[pp.~302, 391--395]{WhittakerWatson80}. Concretely, the Legendre series for $\frac{\omega^2}{c(hc)^3}R(\mu,\mu',\omega)$ reads
\begin{equation}
\label{Legendre_series}
\frac{\omega^2}{c(hc)^3}R(\mu,\mu',\omega) = \frac{1}{4\pi} \sum_{l=0}^{\infty}(2l+1)\phi_l(\omega)\int_{0}^{2\pi}P_l(\cos\theta)d\varphi\, ,
\end{equation}
with the Legendre polynomials $P_l$, $l=0,1,\ldots$, where $\theta$ is the angle between the incoming and the outgoing particle and 
\begin{equation}
\label{cos_in_Legendre}
\cos(\theta) = \mu\mu' + [(1-\mu^2)(1-\mu'^2)]^{\frac{1}{2}}\cos\varphi\,.
\end{equation}

With the first two Legendre polynomials given by
\begin{equation*}
P_0(\cos \theta) = 1\,,\quad
P_1(\cos \theta) = \cos\theta\,, 
\end{equation*}
truncation of the series after the second term provides
\begin{equation*}
\frac{\omega^2}{c(hc)^3}R(\mu,\mu',\omega) \approx \frac{1}{4\pi}\left(\phi_0(\omega)\int_0^{2\pi}1\, d\varphi + 3 \phi_1(\omega)\int_0^{2\pi}\cos(\theta)d\varphi\right)=\frac{1}{2}\phi_0(\omega) + \frac{3}{2}\phi_1(\omega)\mu\mu'\,.
\end{equation*}
Inserting this in the collision integral \eqref{isoscattering}, without explicitly mentioning the dependency on $t$, $r$ and $\omega$, one obtains
\begin{equation}
\mathcal C(f) \approx \frac{1}{2}\int_{-1}^1 (\phi_0 + 3\phi_1\mu\mu')(f(\mu')-f(\mu))d\mu' = -\phi_0 f+\phi_0\frac{1}{2}\int_{-1}^1 fd\mu + 3\mu\phi_1\frac{1}{2}\int_{-1}^1 f\mu d\mu\, ,  \label{coll}
\end{equation}
which is a affine function in $\mu$ expressed in terms of $f$, the zeroth and first angular moments of $f$ and the opacities $\phi_0$ and $\phi_1$. Together with the emissivity and absorptivity of neutrinos by the matter on the right hand side of \eqref{Boltz} the latter gives rise to the definition of the neutrino \emph{mean free path}
\begin{equation}
\label{meanfreepath}
\lambda:= \frac{1}{j+\chi+\phi_0-\phi_1}=\frac{1}{\tilde\chi+\phi_0-\phi_1}\,. 
\end{equation}
This definition is motivated by the fact that $\lambda/3$ occurs as the diffusion parameter in the diffusion limit of the Boltzmann equation that will be derived in the following subsection, see \eqref{equation_compare}. It is clear that the smaller the diffusion parameter is, the smaller the diffusion of neutrinos represented by the diffusion term in \eqref{equation_compare} becomes which physically corresponds to a smaller mean free path.

Finally, we remark that a collision kernel which can be expanded as in \eqref{Legendre_series} with \eqref{cos_in_Legendre} is always symmetric in $\mu$ and $\mu'$ since $\cos(\theta)$ in \eqref{cos_in_Legendre} has this property. For the same reason, \eqref{coll} as well as any truncation of the Legendre series in \eqref{Legendre_series} is symmetric in $\mu$ and $\mu'$.

\subsection{Derivation of the diffusion limit}
\label{diffusion_limit}

Now we outline the line of thought for the derivation of the diffusion limit given in \cite{LiebendoerferEtAl09}. It is based on the truncation of the Legendre expansion of the collision kernel after the second term given in the previous subsection. With \eqref{coll} we obtain the equation
\begin{equation}
\mathcal D(f) = j-(\tilde\chi+\phi_0)f+\phi_0\frac{1}{2}\int_{-1}^1 fd\mu + 3\mu\phi_1\frac{1}{2}\int_{-1}^1 f\mu d\mu\,  \label{Boltz_coll_approx}
\end{equation}
as an approximation of the Boltzmann equation \eqref{Boltz}. The basic idea for the derivation of the diffusion limit is to exploit the special structure of the right hand side in \eqref{Boltz_coll_approx}, i.e., the fact that $f$ can be expressed in terms of $\mathcal D(f)$ and 
the zeroth and first angular moment of~$f$. By taking the zeroth and first angular moments of \eqref{Boltz_coll_approx}, one can therefore express these moments of $f$ in terms of the zeroth and first angular moments of $\mathcal D(f)$, i.e., eliminate them in \eqref{Boltz_coll_approx} and thus express $f$ solely in terms of $\mathcal D(f)$ and its zeroth and first angular moment.

Concretely, since $j$ is isotropic and the collision kernel in \eqref{coll} that appears in the right hand side of \eqref{Boltz_coll_approx} is symmetric in $\mu$ and $\mu'$, i.e., \eqref{collision_vanishes} is applicable, the zeroth moment of \eqref{Boltz_coll_approx} reduces to
\begin{equation}\label{Boltz_coll_approx_mean}
\frac{1}{2}\int_{-1}^1 \mathcal D(f) d\mu = j - \tilde \chi \frac{1}{2}\int_{-1}^1 f d\mu\,,
\end{equation}
i.e.,
\begin{equation}
\label{D1}
\frac{1}{2}\int_{-1}^1 f d\mu  = \frac{1}{\tilde \chi}\left(j-  \frac{1}{2}\int_{-1}^1 \mathcal D(f) d\mu\right)\,.
\end{equation}

In the first moment of \eqref{Boltz_coll_approx}, the first and third summand on the right hand side vanish since they are independent of $\mu$ and the first angular moment of a constant is zero. Considering $\frac{1}{2}\int_{-1}^1 3\mu^2d\mu=1$, the coefficient in the last summand reduces to $\phi_1$ so that we obtain
\begin{equation*}
\frac{1}{2}\int_{-1}^1 \mathcal D(f) \mu d\mu = -(\tilde \chi+\phi_0-\phi_1)\frac{1}{2}\int_{-1}^1f \mu d\mu\, ,
\end{equation*}
which, by \eqref{meanfreepath}, leads to 
\begin{equation}
\label{D2}
\frac{1}{2}\int_{-1}^1f \mu d\mu  = -\lambda\frac{1}{2}\int_{-1}^1 \mathcal D(f) \mu d\mu\,.
\end{equation}
With \eqref{D1} and \eqref{D2} on the right hand side of \eqref{Boltz_coll_approx} one can now express $f$ in terms of $\mathcal D(f)$ and its first two moments as
\begin{equation}
\label{f_inDf}
f = \frac{1}{\tilde \chi + \phi_0}\left[ j -\mathcal D(f) + \frac{\phi_0}{\tilde \chi}\left(j -\frac{1}{2}\int_{-1}^{1}\mathcal D(f)d\mu \right) - 
3\mu\phi_1\lambda\frac{1}{2}\int_{-1}^1 \mathcal D(f) \mu d\mu \right]\,.
\end{equation}

Now a Chapman--Enskog--like expansion is performed in \cite{LiebendoerferEtAl09}. Therefore, a small parameter $\eps$ is introduced in order to scale the emissivity and opacity terms
$$
j=\frac{\bar j}{\varepsilon}\,, \quad
\tilde\chi= \frac{\bar{\tilde{\chi}}}{\varepsilon}\,, \quad
\phi_0= \frac{\bar \phi_0}{\varepsilon}\,,\quad
\phi_1= \frac{\bar \phi_1}{\varepsilon}\, ,
$$
which are considered to be large. Thus the expansion is performed for small mean free paths $\lambda=\eps\bar\lambda$ with $\bar\lambda=(\bar{\tilde{\chi}}+\bar\phi_0-\bar\phi_1)^{-1}$. Inserting this scaling in \eqref{f_inDf} leads to
\begin{equation}
\label{f_inDf_eps}
f = \frac{\eps}{\bar{\tilde{\chi}} + \bar\phi_0}\left[ \frac{\bar j}{\eps} -\mathcal D(f) + \frac{\phi_0}{\tilde \chi}\left(\frac{\bar j}{\eps} -\frac{1}{2}\int_{-1}^{1}\mathcal D(f)d\mu \right) - 
3\mu\phi_1\lambda\frac{1}{2}\int_{-1}^1 \mathcal D(f) \mu d\mu \right]\,.
\end{equation}
If we expand $f=f_0+\eps f_1+\eps^2 f_2+\ldots$ and collect the terms of the same power of $\eps$ in \eqref{f_inDf_eps}, using the linearity (and formally the continuity) of $\mathcal{D}$, we obtain the zeroth order term
\begin{equation*}
f_0=\frac{j}{{\tilde{\chi}} + \phi_0}\left(1+\frac{\phi_0}{\tilde \chi}\right)=\frac{j}{\tilde{\chi}}\, ,
\end{equation*}
and the first order term
\begin{equation*}
\varepsilon f_1 = \frac{-1}{\tilde \chi+\phi_0}\left[ \mathcal D(f_{0})+ \frac{\phi_0}{\tilde \chi}\frac{1}{2}\int_{-1}^{1}\mathcal D(f_0)d\mu +  3\mu\phi_1\lambda\frac{1}{2}\int_{1}^1 \mathcal D(f_0) \mu d\mu \right]\,.
\end{equation*}

These expressions are now used to compute the approximation 
$$
s=\frac{1}{2}\int_{-1}^1 \mathcal D(f_0+ \varepsilon f_1) d\mu\, ,
$$
the angularly integrated left hand side of \eqref{Boltz_coll_approx}, which by \eqref{Boltz_coll_approx_mean} is equal to {{{}}$j-\tilde\chi\, \dfrac{1}{2}\displaystyle\int_{-1}^1 fd\mu$}\,, the so-called particle number exchange rate with matter or total interaction rate. In order to simplify the computation, it is helpful to decompose the operator $\mathcal D$ additively into a part $\mathcal D^+$ that is symmetric with respect to $\mu$ and a part $\mathcal D^-$ that is antisymmetric with respect to $\mu$, i.e., we write
 \begin{equation*}
 \mathcal D(f) = \mathcal D^+(f) + \mathcal D^-(f)\,,
 \end{equation*}
with
 \begin{equation}
 \label{D+}
 \mathcal D^+(f) = \frac{d f}{c dt}  + \left[ \mu \left( \frac{d \ln \rho}{cd t} + \frac{3v}{cr} \right) \right] (1-\mu^2)\frac{\partial f}{\partial \mu} + \left[ \mu^2 \left( \frac{d \ln \rho}{cd t} + \frac{3v}{cr} \right) - \frac{v}{cr}\right] \omega \frac{\partial f}{\partial \omega}\, ,
 \end{equation}
and 
 \begin{equation}
 \label{D-}
 \mathcal D^-(f) = \mu \frac{\partial f}{\partial r} + \frac{1}{r}(1-\mu^2)\frac{\partial f}{\partial \mu}\,.
 \end{equation}

Calculating
\begin{equation}
\label{s}
s = \frac{1}{2}\int_{-1}^1 \mathcal D(f_0+ \varepsilon f_1) d\mu = \frac{1}{2}\int_{-1}^1 \mathcal D^+(f_0)d\mu + \frac{1}{2}\int_{-1}^1 \mathcal D^-(f_0)d\mu+ \frac{1}{2}\int_{-1}^1\mathcal D^+(\varepsilon f_1)d\mu+\frac{1}{2}\int_{-1}^1\mathcal D^-(\varepsilon f_1)d\mu\, ,
\end{equation}
we immediately see $\frac{1}{2}\int_{-1}^1 \mathcal D^-(f_0) d\mu = 0$ since $\mathcal D^-$ is antisymmetric in $\mu$ and $f_0$ does not depend on $\mu$. For the third summand we obtain
\begin{multline*}
\frac{1}{2}\int_{-1}^1 \mathcal D^+(\varepsilon f_1) d\mu
=\\
\frac{1}{2}\int_{-1}^1 \mathcal D^+\Bigg(\frac{-1}{\tilde \chi+\phi_0}\Bigg[ \mathcal D^+(f_{0})+\mathcal D^-(f_0)
+\frac{\phi_0}{\tilde \chi}\frac{1}{2}\int_{-1}^{1}(\mathcal D^+(f_0)+\mathcal D^-(f_0))d\mu
+3\mu\phi_1\lambda\frac{1}{2}\int_{-1}^1 (\mathcal D^+(f_0)+\mathcal D^-(f_0)) \mu d\mu \Bigg]\Bigg) d\mu\,.
\end{multline*}
Since $f_0$ does not depend on $\mu$, the term $\mathcal D^+(f_0)\mu$ in the second inner integral of this expression is an antisymmetric polynomial in $\mu$ and thus vanishes after integration. By the same reasoning, the term $\mathcal D^-(f_0)\mu$ in the second inner integral is a symmetric polynomial in $\mu$ so that angular integration gives an expression that no longer depends on $\mu$. Consequently, the last summand in the brackets is linear in $\mu$ so that the application of the operator $\mathcal D^+(\frac{-1}{\tilde \chi+\phi_0}(\cdot))$ to this expression is an antisymmetric polynomial in $\mu$ that vanishes after the outer integration. By the same reasoning, the second summand in the brackets vanishes after the application of this operator and the outer integration. With these arguments and $\frac{1}{2}\int_{-1}^1 \mathcal D^-(f_0) d\mu = 0$ as seen above we obtain the simplified equation
\begin{equation}
\label{D+neglect}
\frac{1}{2}\int_{-1}^1 \mathcal D^+(\varepsilon f_1) d\mu
=
\frac{1}{2}\int_{-1}^1 \mathcal D^+\Bigg(\frac{-1}{\tilde \chi+\phi_0}\Bigg[ \mathcal D^+(f_{0})
+\frac{\phi_0}{\tilde \chi}\frac{1}{2}\int_{-1}^{1}\mathcal D^+(f_0)d\mu\Bigg]\Bigg) d\mu\, ,
\end{equation}
in which the right hand side only contains terms that undergo the application of the operator $\mathcal D^+$ twice. In \cite{LiebendoerferEtAl09} these expressions are neglected in the calculation of $s$ because they are considered ``of higher order'' than the ``leading order term'' $\frac{1}{2}\int_{-1}^1 \mathcal D^+(f_0)d\mu$ whose contribution is already considered in \eqref{s}.

The last summand in \eqref{s} is given by
\begin{multline*}
\frac{1}{2}\int_{-1}^1 \mathcal D^-(\varepsilon f_1) d\mu
=\\
\frac{1}{2}\int_{-1}^1 \mathcal D^-\Bigg(\frac{-1}{\tilde \chi+\phi_0}\Bigg[ \mathcal D^+(f_{0})+\mathcal D^-(f_0)
+\frac{\phi_0}{\tilde \chi}\frac{1}{2}\int_{-1}^{1}(\mathcal D^+(f_0)+\mathcal D^-(f_0))d\mu
+3\mu\phi_1\lambda\frac{1}{2}\int_{-1}^1 (\mathcal D^+(f_0)+\mathcal D^-(f_0)) \mu d\mu \Bigg]\Bigg) d\mu\,.
\end{multline*}
As for the third summand, the term $\mathcal D^+(f_0)\mu$ vanishes after integration in the second inner integral. Since $f_0$ is independent of $\mu$, the term $\mathcal D^+(f_0)$, in the first inner integral is a symmetric polynomial in $\mu$. Therefore, since $\frac{1}{2}\int_{-1}^1 \mathcal D^-(f_0) d\mu = 0$ the first inner integral does no longer depend on $\mu$ so that the application of the operator $\mathcal D^-(\frac{-1}{\tilde \chi+\phi_0}(\cdot))$ to it is linear in $\mu$ and vanishes after the outer integration. The same holds for the first term $\mathcal D^+(f_0)$ in the brackets which is a symmetric polynomial in $\mu$ so that the application of $\mathcal D^-(\frac{-1}{\tilde \chi+\phi_0}(\cdot))$ to it gives an antisymmetric polynomial in $\mu$ that vanishes by the outer integration. Consequently, we obtain the equation
\begin{equation*}
\frac{1}{2}\int_{-1}^1 \mathcal D^-(\varepsilon f_1) d\mu=
\frac{1}{2}\int_{-1}^1 \mathcal D^-\Bigg(\frac{-1}{\tilde \chi+\phi_0}\Bigg[ \mathcal D^-(f_{0})
+3\mu\phi_1\lambda\frac{1}{2}\int_{-1}^1 \mathcal D^-(f_0) \mu d\mu\Bigg]\Bigg) d\mu\, ,
\end{equation*}
that we can further simplify by observing 
\begin{equation}
\label{D-f0}
\mathcal D^-(f_0)=\mu\frac{\partial f_0}{\partial r}\,,
\end{equation}
which leads to
$$ 
3\mu\phi_1\lambda\frac{1}{2}\int_{-1}^1 \mathcal D^-(f_0) \mu d\mu=3\mu\phi_1\lambda\frac{1}{2}\int_{-1}^1 \mu^2\frac{\partial f_0}{\partial r} d\mu=\phi_1\lambda\mu\frac{\partial f_0}{\partial r}=\phi_1\lambda\mathcal D^-(f_0)\,,
$$
so that, with \eqref{meanfreepath}, we can conclude
$$
\frac{1}{2}\int_{-1}^1 \mathcal D^-(\varepsilon f_1) d\mu=
\frac{1}{2}\int_{-1}^1 \mathcal D^-\Bigg(\frac{-1}{\tilde \chi+\phi_0}\Big[ \big( 1
 +\phi_1\lambda \big)\mathcal D^-(f_0)\Big]\Bigg) d\mu
 = \frac{1}{2}\int_{-1}^1 \mathcal D^-\left(-\lambda\mathcal D^- (f_0)\right)d\mu\,.
$$
Altogether we obtain the approximation
\begin{equation}
\label{s_leading}
s=
\frac{1}{2}\int_{-1}^1 \mathcal D^+(f_0) d\mu
-
\frac{1}{2}\int_{-1}^1 \mathcal D^-\left(\lambda\mathcal D^- (f_0)\right)d\mu\, ,
\end{equation}
of the total interaction rate $s$ ``in leading order''. Note that there is no contribution to $s$ in \eqref{s} that involves only one application of the operator $\mathcal D^-$.

To determine this approximation for $s$ in concrete terms, we insert \eqref{D+} and \eqref{D-} in \eqref{s_leading}. First, we calculate
\begin{align*}
\frac{1}{2}\int_{-1}^1 \mathcal D^+(f_0) d\mu 
&=  \frac{1}{2}\int_{-1}^1\left[ \frac{d f_0}{c dt}  + \left[ \mu \left( \frac{d \ln \rho}{cd t} + \frac{3v}{cr} \right) \right] (1-\mu^2)\frac{\partial f_0}{\partial \mu} + \left[ \mu^2 \left( \frac{d \ln \rho}{cd t} + \frac{3v}{cr} \right) - \frac{v}{cr}\right] \omega \frac{\partial f_0}{\partial \omega}\right] d\mu\\[2mm]
&= \frac{d f_0}{c dt} + \left[ \frac{1}{3} \left( \frac{d \ln \rho}{cd t} + \frac{3v}{cr} \right) -\frac{v}{cr} \right]\omega \frac{\partial f_0}{\partial \omega}\\[2mm]
&= \frac{d f_0}{c dt} +  \frac{1}{3} \frac{d \ln \rho}{cd t}  \omega \frac{\partial f_0}{\partial \omega}\,.
\end{align*}
Here, the second line is obtained by interchanging integration and differentiation, considering the notation \eqref{f_t_is_beta_t} for $f_0$ in the first and third summand in the integral, using $\frac{1}{2}\int_{-1}^1 \mu^2d\mu=\frac{1}{3}$ for the latter and observing that the second summand vanishes since $f_0$ does not depend on $\mu$. Recall that the same reasoning already led to the left hand side of the trapped particle equation \eqref{eq:trapped}. Therefore, here, the second term on the right hand side of~\eqref{s_leading} is the more interesting one that will lead to the definition of the diffusion source. Concretely, taking \eqref{D-f0} into account, we see
\begin{align*}
\frac{1}{2}\int_{-1}^1 \mathcal D^-\left(\lambda\mathcal D^- (f_0)\right)d\mu 
&= \frac{1}{2}\int_{-1}^1 \mathcal D^-\left(\lambda\mu\frac{\partial f_0}{\partial r}\right)d\mu\\[2mm]
&= \frac{1}{2}\int_{-1}^1\left[ \mu \frac{\partial}{\partial r}  \left(\lambda\mu\frac{\partial f_0}{\partial r}\right)  + \frac{1}{r}(1-\mu^2)\frac{\partial }{\partial \mu} \left(\lambda\mu\frac{\partial f_0}{\partial r}\right) \right]     d\mu\\[2mm]
&= \frac{1}{3} \frac{\partial}{\partial r}\left(  \lambda\frac{\partial f_0}{\partial r}\right) +  \frac{1}{2}\int_{-1}^1  \frac{1}{r}(1-\mu^2) \lambda\frac{\partial f_0}{\partial r} d\mu\\[2mm]
&= \frac{1}{3} \frac{\partial}{\partial r}\left(  \lambda\frac{\partial f_0}{\partial r}\right) + \frac{2}{3}\frac{1}{r}\lambda \frac{\partial f_0}{\partial r}\\[2mm]
&= \frac{1}{r^2}\frac{\partial }{\partial r}\left( r^2\frac{\lambda}{3}\frac{\partial f_0}{\partial r}\right)\,.
\end{align*}
For the third and fourth line we used again $\frac{1}{2}\int_{-1}^1 \mu^2d\mu=\frac{1}{3}$, the isotropy of $f_0$ and $\lambda$ as well as the notation as in \eqref{f_t_is_beta_t} for $f_0$. The last line follows from the product rule. As a result, considering \eqref{nabla_div_in_r}, we obtain a diffusion term induced by $f_0$ with $\lambda/3$
as the (small!) diffusion parameter.

Finally, for the derivation of $\Sigma$ in \eqref{decomposedtrapped} and \eqref{decomposedstreaming} we consider $f=f_0+\eps f_1$ in the Boltzmann equation \eqref{Boltz_pure} with the approximated collision integral as in \eqref{Boltz_coll_approx} and set $f^t=f_0$ and $f^s=\eps f_1$. Then, a first order approximation in $\eps$ of the angularly integrated \eqref{Boltz_pure} or \eqref{Boltz_coll_approx} gives
\begin{equation}
\label{equation_compare}
\frac{d f^t}{c dt} +  \frac{1}{3} \frac{d \ln \rho}{cd t}  \omega \frac{\partial f^t}{\partial \omega}-\frac{1}{r^2}\frac{\partial }{\partial r}\left( r^2\frac{\lambda}{3}\frac{\partial f^t}{\partial r}\right)
=j-\tilde\chi \left(f^t+\frac{1}{2}\int_{-1}^1f^sd\mu\right)\,,
\end{equation}
``in leading order'', i.e., neglecting the terms in \eqref{D+neglect}. On the right hand side we use again \eqref{f_t_is_beta_t} as well as~\eqref{collision_vanishes}. Note that $f^s$ does not need to be isotropic. Now, if we compare equation \eqref{equation_compare} with the trapped particle equation \eqref{eq:trapped} we obtain
\begin{equation}
\label{Sigma_diffusion}
\Sigma_{\mbox{\tiny diff}}:=-\frac{1}{r^2}\frac{\partial }{\partial r}\left( r^2\frac{\lambda}{3}\frac{\partial f^t}{\partial r}\right)+\frac{\tilde\chi}{2}\int_{-1}^1f^sd\mu\, ,
\end{equation}
as a suitable definition for $\Sigma$ in this limit. Since here, the first term is a diffusion term, equation \eqref{equation_compare} can be regarded as an approximation of the Boltzmann equation in the diffusion limit. Therefore, we call $\Sigma$ a diffusion source. Observe from the above calculations that the diffusion term on the left hand side of \eqref{equation_compare} stems from the contribution of $\frac{1}{2}\int_{-1}^1 \mathcal D(f^s) d\mu$ to the total interaction rate ``in leading order''. We announce that in a forthcoming paper \cite{BeFrGaLiMiVa12}, the diffusion limit will be derived ``in leading order'' by a Chapman--Enskog expansion and, even without the ``leading order'' approximation, by a Hilbert expansion of the Boltzmann equation.

The three coupled equations \eqref{eq:trapped}, \eqref{eq:streaming} and \eqref{Sigma_diffusion} arise from taking the angular mean of \eqref{decomposedtrapped} and \eqref{decomposedstreaming} and \eqref{Boltz_coll_approx}, i.e., they are no longer dependent on $\mu$. However, in spite of \eqref{collision_vanishes}, the influence of the collision integral is contained ``in leading order'' in $\Sigma\left(\omega,f^t,\frac{1}{2}\int_{-1}^1 f^s d\mu,j,\mathcal J\right)$
by its dependency on the mean free path $\lambda=(\tilde\chi+\phi_0-\phi_1)^{-1}$.

If the mean free path is too big, the diffusion limit \eqref{equation_compare} is no longer a good approximation of the Boltzmann equation \eqref{Boltz} so that the diffusion term in \eqref{Sigma_diffusion} may become too big. Concretely, if we neglect the dynamics of the background matter, i.e., the second term on the left hand side in \eqref{eq:trapped}, then $f^t$ tends exponentially towards the stationary state solution 
\begin{equation}
\label{stationary_state}
f_\infty^t=\frac{j-\Sigma}{\tilde\chi}\,, 
\end{equation}
compare \eqref{analytical_solution}. Since $f_\infty^t\geq 0$ is an a priori condition for a distribution function, we obtain the necessary condition $\Sigma\leq j$ for the coupling term. In \cite[p.\,1177]{LiebendoerferEtAl09} this condition is obtained by physical arguments for large mean free paths where the streaming particle component $f^s$ dominates over the trapped particle component~$f^t$. In \cite{BeFrGaLiMiVa12} the \emph{free streaming limit} $\Sigma=j$ will be derived by a Hilbert expansion of the Boltzmann equation.

Conversely, if the mean free path $\lambda$ tends to $0$, i.e., $\eps\to 0$, then $f^t$ dominates over $f^s=\eps f_1$. In the limit $\eps=0$ we obtain $\Sigma=0$ in \eqref{Sigma_diffusion}. By \eqref{stationary_state} and the definition of $\tilde\chi=j+\chi$, we have at least the requirement $\Sigma\geq-\chi$ since the distribution function always satisfies $f_\infty^t\leq 1$. In \cite[p.\,1177]{LiebendoerferEtAl09} it is argued physically that $f_\infty^t$ should be at most $j/\tilde\chi$, which was the zeroth order approximation of $f$ above, because that function represents the distribution for thermal equilibrium. With this assumption one gets $\Sigma\geq 0$ from \eqref{stationary_state}. In \cite{BeFrGaLiMiVa12} the \emph{reaction limit} $\Sigma=0$ will also be derived by a Hilbert expansion of the Boltzmann equation.

With these considerations regarding the free streaming and the reaction limits, the diffusion source in \eqref{Sigma_diffusion} is limited from above by $j$ and below by $0$ in order to account for regimes where the diffusion limit is not valid. Altogether, with
\begin{equation}
\label{Sigma}
\Sigma := \min \left\{\max\left[-\frac{1}{r^2}\frac{\partial}{\partial r}\left(r^2\frac{\lambda}{3}\frac{\partial f^t}{\partial r}\right) + \frac{\tilde\chi}{2}\int_{-1}^1f^s d\mu\,,\ 0\,\right],\ j \right\}\, ,
\end{equation}
as the coupling term, the equations \eqref{eq:trapped} and \eqref{eq:streaming} with \eqref{f_t_is_beta_t} and \eqref{scattering_sphere} give the Isotropic Diffusion Source Approximation (IDSA) of the Boltzmann equation as introduced in \cite{LiebendoerferEtAl09}.

\section{Reduced Boltzmann model problem}
\label{reducedSec}

In this section, we reduce the Boltzmann equation \eqref{Boltz} to a simpler one that will later serve as a model to perform first numerical tests.
\subsection{Assumptions}
For our reduction we decouple the Boltzmann equation \eqref{Boltz} from the state of the matter that contributes to angular aberration $F_\mu^1\frac{\partial f}{\partial\mu}$, Doppler shift $F_\omega\frac{\partial f}{\partial\omega}$, emissivity $j$, opacity $\tilde\chi$ and scattering kernel $R$. The first assumption that we make is that the matter is frozen, so that its state does not depend on time, which implies $F_\mu^1 = F_\omega = 0 $. We also assume that the emissivity and the opacity only depend on $r$, i.e, $j(r)$ and $\tilde\chi(r)$. The third hypothesis is that there are no collisions and, therefore, $R = 0$ and $\mathcal C(f) = 0$. Under these hypotheses the Boltzmann equation \eqref{Boltz} reduces to 
\begin{equation}\label{BoltzRed0}
\frac{1}{c}\frac{\partial f(t,r,\mu,\omega)}{\partial t}+\mu \frac{\partial f(t,r,\mu,\omega)}{\partial r} + \frac{1}{r}(1-\mu^2)\frac{\partial f(t,r,\mu,\omega)}{\partial \mu} = j(r)-\tilde\chi(r) f(t,r,\mu,\omega)\,.
\end{equation}

For simplicity, we rescale the time $t' = ct$ and no longer mention the dependency on $\omega$ from now on since equation \eqref{BoltzRed0} is monochromatic, i.e., $\omega$ does only appear as a parameter here. Dropping the prime in the time scaling again, the reduced equation that we want to solve is
\begin{equation}\label{BoltzRed1}
\frac{\partial f(t,r,\mu)}{\partial t}+\mu \frac{\partial f(t,r,\mu)}{\partial r} + \frac{1}{r}(1-\mu^2)\frac{\partial f(t,r,\mu)}{\partial \mu} = j(r)-\tilde\chi(r) f(t,r,\mu)\,.
\end{equation}

\subsection{Conservative formulation of transport equation}

In this subsection, we will show that equation \eqref{BoltzRed1} can be written in conservative form as 
\begin{equation}\label{conserv}
\frac{\partial f}{\partial t} +\nabla\cdot \left(\begin{pmatrix}\mu\\ \frac{1}{r}(1-\mu^2)\end{pmatrix}f\right) = j -\tilde\chi f\,,
\end{equation}
where the divergence operator is given by
$
\nabla\cdot\begin{pmatrix}a\\b\end{pmatrix}=\frac{1}{r^2}\frac{\partial }{\partial r} r^2 a + \frac{\partial}{\partial \mu}b\,.
$
\begin{lmm} With the divergence operator just mentioned, we have 
$
\nabla \cdot \begin{pmatrix}\mu\\ \frac{1}{r}(1-\mu^2)\end{pmatrix} = 0.
$
\end{lmm}
{\bf Proof} By direct computation we calculate
$
\nabla\cdot\begin{pmatrix}\mu\\\frac{1}{r}(1-\mu^2)\end{pmatrix}=\frac{1}{r^2}\frac{\partial }{\partial r} r^2 \mu + \frac{1}{r}\frac{\partial}{\partial \mu}(1-\mu^2) = \frac{2\mu}{r}-\frac{2\mu}{r} = 0.
$

\begin{rmrk} The divergence operator we use can be derived from the 6D Cartesian divergence operator in phase space if we assume spherical symmetry and constant velocity. The radial term in the divergence operator is the usual radial term in spherical symmetry and the additional term represents the term for the velocity.
\end{rmrk}

\subsection{Reduced form of the IDSA equations}
In order to compare the reduced Boltzmann equation with the equations of IDSA, we need to apply the same assumptions as above. The reduced system of equations corresponding to \eqref{eq:trapped}, \eqref{eq:streaming} and \eqref{Sigma} turns out to be the following.
\begin{equation}
\label{eq:trappedR}
\frac{d f^t}{cd t}  = j-\tilde \chi f^t -\Sigma\, ,
\end{equation}
\begin{equation}
\label{eq:streamingR}
\frac{1}{r^2}\frac{\partial}{\partial r}\left( \frac{r^2}{2}\int_{-1}^1f^s\mu d\mu\right) = -\frac{\tilde\chi}{2}\int_{-1}^1f^s d\mu+\Sigma\, ,
\end{equation}
\begin{equation*}
\Sigma = \min \left\{\max\left[-\frac{1}{r^2}\frac{\partial}{\partial r}\left(r^2\frac{\lambda}{3}\frac{\partial f^t}{\partial r}\right) + \frac{\tilde\chi}{2}\int_{-1}^1f^s d\mu\, ,\ 0\,\right],\ j \right\}\, .
\end{equation*}
\vspace{0.1cm}

The only changes compared with \eqref{eq:trapped}, \eqref{eq:streaming} and \eqref{Sigma} are that the Doppler shift term of the trapped particle equation is absent and that the definition of the mean free path now is $\lambda := \tilde\chi^{-1}$.

\section{Numerical treatment}
\label{numericaltreatSec}

\subsection{Grid}
The spherically symmetric computational domain for the IDSA, $\Omega_{\mbox{\tiny IDSA}}^T = [0,R]\times[0,T]$, and the one for the Boltzmann model, $\Omega_{\mbox{\tiny Bol}}^T = [0,R]\times[-1,1]\times[0,T]$, is represented by points $(r_i, t^n)$ and $(r_i, \mu_j, t^n)$, respectively. We fix $i_{\max}$, $j_{\max}$ and $n_{\max}$ and define the grid by 
$$r_{i+1/2} = \frac{iR}{i_{\max}}\,,\quad \mu_{j+1/2} = -1 +\frac{2j}{j_{\max}}\,,\quad t^n = \frac{nT}{n_{\max}}\,.$$
We also define the cells $C_{ij}$ by
$$
C_{ij} = [r_{i-1/2},r_{i+1/2}]\times[\mu_{j-1/2},\mu_{j+1/2}]\,.
$$

\subsection{Reduced model}
\subsubsection{Time splitting}
In order to solve \eqref{conserv}, we perform an order one \emph{time splitting} by denoting
\begin{equation*}
F_1(f):=j-\tilde\chi f\quad\mbox{and}\quad F_2(f):=\nabla\cdot \left(\begin{pmatrix}\mu\\
\frac{1}{r}(1-\mu^2)\end{pmatrix}f\right)\,,
\end{equation*}
and writing equation \eqref{conserv}, which is an autonomous ODE with respect to $f$, in the form
\begin{equation}
\label{conserv2}
\frac{\partial f}{\partial t} = F(f) := F_1(f)+F_2(f)\,.
\end{equation}
Denoting the flow maps of the vector fields $F$, $F_1$ and $F_2$ by $\Phi_{t,F}$, $\Phi_{t,F_1}$ and $\Phi_{t,F_2}$, we approximate $\Phi_{t,F}$ by $\Psi_{t,F} :=  \Phi_{t,F_2} \circ \Phi_{t,F_1}$. This is known as a Lie--Trotter splitting and gives an approximation of order $1$ to the solution of \eqref{conserv2}, see \cite[p.\,42]{HairerLubichWanner02}.

\subsubsection{Analytical treatment of the reaction term}\label{AnaReac}

The flow map $\Phi_{t,F_1}$ corresponds to the ODE
\begin{equation*}
\frac{\partial f}{\partial t}= j-\tilde\chi f\, ,
\end{equation*}
that has the {\emph{analytical solution}}
\begin{equation}
\label{analytical_solution}
f(t) = f(0)e^{-t\tilde\chi} + (1-e^{-t\tilde\chi})\frac{j}{\tilde\chi}\, ,
\end{equation}
for every $r\in [0,R]$, $\mu\in [-1,1]$ and $\omega\in [0,E]$.
It describes the exponential change from $f(0)$ to the distribution function $j/{\tilde\chi}$ 
which is known to be a thermal equilibrium function and, therefore, a Fermi--Dirac distribution
$$
\frac{j}{\tilde\chi}=\frac{1}{e^{\frac{\omega-\gamma}{k\vartheta}}+1}\, ,
$$
because of the fermionic nature of the neutrinos \cite[pp.\,1177/9]{LiebendoerferEtAl09}. In the Fermi--Dirac distribution, $\gamma$ is the chemical potential, $k$ is Boltzmann's constant and $\vartheta$ is the temperature of the matter.

\subsubsection{Treatment of the transport term}\label{flow}

\begin{figure}[h!]
\label{neutrino_flux}
\centering
\vspace{-0.2cm}
\includegraphics[width = 0.35\linewidth]{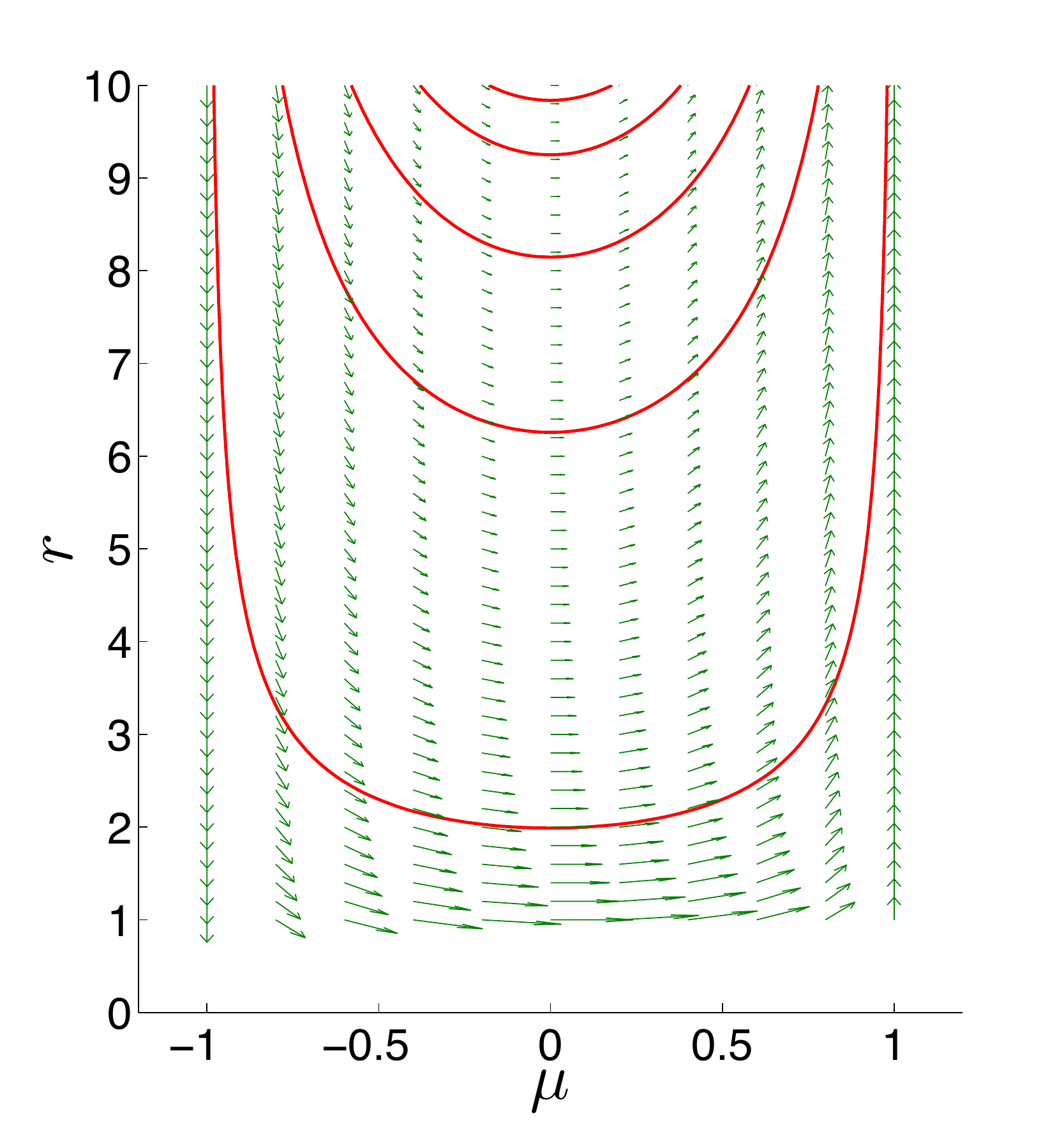}
\vspace{-3mm}\caption{Flow map of the flux of neutrinos due to transport and without reactions. We can see that incoming neutrinos ($\mu<0$) turn into outgoing neutrinos ($\mu>0$) and, therefore, eventually leave the computational domain again.}
\label{flowGraph}
\end{figure}

Figure \ref{flowGraph} shows the behaviour of the flow map $\Phi_{t,F_2}$ in the domain $(r, \mu)\in\Omega = [0,10]\times[-1,1]$.

\begin{rmrk}\label{FlowCFL}
Figure \ref{flowGraph} shows that the flow becomes large as the radius goes to zero. As a consequence, the CFL condition will be critical in the region where $r$ is small if the overall flux $F(f)$ is dominated by $F_2(f)$. On the other hand, if in this region the overall flux is dominated by $F_1(f)$, then the exponential decrease towards equilibrium removes the dependency on $\mu$ for small $r$ as $f$ becomes isotropic.
\end{rmrk}

We now derive a finite volume scheme for the transport term. The \emph{divergence theorem} applied in our coordinates is given by
\begin{equation*}
\int_\Omega \nabla\cdot\begin{pmatrix}a\\b\end{pmatrix} r^2 dr d\mu = \int_{\partial \Omega}\begin{pmatrix}a\\b\end{pmatrix}\cdot \vec n\ T_{\partial \Omega}(r^2 dr d\mu)\, ,
\end{equation*}
with $T_{\partial \Omega}(r^2 dr d\mu)$ denoting the trace measure.

Integrating \eqref{conserv} on a cell $C_{ij}$ and defining $f^n_{ij} :=\frac{1}{|C_{ij}|} \int_{C_{ij}}f(t_n,r,\mu) r^2 dr d\mu$ with a forward Euler scheme, we obtain the \emph{finite volume scheme} 
\begin{multline*}
f_{ij}^{n+1} = f_{ij}^n -\Delta t\left[\int_{\mu_{j-1/2}}^{\mu_{j+1/2}}-\mu f_{ij}^n r^2_{i-1/2} d\mu  + \int_{r_{i-1/2}}^{r_{i+1/2}}(1-\mu^2_{j+1/2})f_{ij}^n r dr\right.\\
\left. +\int_{\mu_{j+1/2}}^{\mu_{j-1/2}} \mu f_{i+1,j}^n r^2_{i+1/2} d\mu  - \int_{r_{i+1/2}}^{r_{i-1/2}}(1-\mu^2_{j-1/2})f_{i,j-1}^n r dr\right]\,,
\end{multline*}
for $\mu<0$ and
\begin{multline*}
f_{ij}^{n+1} = f_{ij}^n -\Delta t\left[\int_{\mu_{j-1/2}}^{\mu_{j+1/2}}-\mu f_{i-1,j}^n r^2_{i-1/2} d\mu  + \int_{r_{i-1/2}}^{r_{i+1/2}}(1-\mu^2_{j+1/2})f_{ij}^n r dr\right.\\
\left. +\int_{\mu_{j+1/2}}^{\mu_{j-1/2}} \mu f_{ij}^n r^2_{i+1/2} d\mu  - \int_{r_{i+1/2}}^{r_{i-1/2}}(1-\mu^2_{j-1/2})f_{i,j-1}^n r dr\right]\,,
\end{multline*}
for $\mu >0$. This scheme uses a first order upwind approximation of the flux. 

\subsection{Implicit-explicit finite difference discretization of the IDSA}

In this section, we outline the discretization of the IDSA as introduced in \cite[App.~B]{LiebendoerferEtAl09} and adapt it to our case.

\subsubsection{Computation of the mean of the streaming particle}

The time integration of IDSA is done by taking equation \eqref{eq:streamingR}, i.e.,
$$\frac{1}{r^2}\frac{\partial}{\partial r}\left(r^2\frac{1}{2}\int_{-1}^1 f^s \mu d\mu \right) = \Sigma-\tilde\chi \frac{1 }{2}\int_{-1}^1 f^s d \mu\, ,\quad$$
and integrating it over space
\begin{equation}\label{eq_streaming_D}
\frac{1}{2}\int_{-1}^1 f^s \mu d\mu  = \frac{1}{r^2}\int_0^r \left[\Sigma-\tilde\chi \frac{1 }{2}\int_{-1}^1 f^s d \mu \right]r^2 dr\,.
\end{equation}
There is no explicit time dependence in this equation as the streaming component is in the stationary state limit, see Subsection \ref{Hyp}. Hence, time integration of this component reduces to updating the angular mean of~$f^s$.
Taking the zeroth and first moments of $f^s$ as variables, we discretize equation \eqref{eq_streaming_D} explicitly by 
\begin{equation*}
\left(\frac{1}{2}\int_{-1}^1 f^s \mu d\mu \right)_{i+\frac{1}{2}}^{n+1} = \frac{1}{r_{i+\frac{1}{2}}^2}\sum_{j=1}^{j=i+\frac{1}{2}}\left[ \Sigma_j^n -\tilde\chi_j^n\left(\frac{1}{2}\int_{-1}^1 f^s d\mu \right)_j^n \right]r_j^2(r_{j+\frac{1}{2}}-r_{j-\frac{1}{2}})\,.
\end{equation*}
This is a mid-point rule formula. Once we have the flux $\frac{1}{2}\int_{-1}^1f^s \mu d\mu$, we compute the updated $\frac{1}{2}\int_{-1}^1 f^s d\mu$ by discretizing \eqref{scattering_sphere} as
\begin{equation*}
\left(\frac{1}{2}\int_{-1}^1 f^s d\mu \right)_{i}^{n+1} = \frac{2}{1+\sqrt{1-\left(\frac{R_\nu^n}{\max(r_i,R_\nu^n)} \right)^2}}\left(\frac{r_{i-\frac{1}{2}}}{r_i} \right)^2 \max\left[ \left(\frac{1}{2}\int_{-1}^1 f^s \mu d\mu \right)_{i-\frac{1}{2}}^{n+1},\,0 \right]\,.
\end{equation*}
As in \cite[p.~1188]{LiebendoerferEtAl09}, we use the factor $\left(\frac{r_{i-\frac{1}{2}}}{r_i} \right)^2$ to convert the flux from the inner zone edge $i-\frac{1}{2}$, where it has been computed, to the zone center $i$, where it is used in the computation of the diffusion source. The maximum operator forces the flux not to be directed against the density gradient. In \cite[p.~1188]{LiebendoerferEtAl09}, it is claimed that this increases the stability of the scheme. We did not test this statement here. The value of the neutrino scattering sphere radius $R_{\nu}^n$ is considered as given in our model.

\subsubsection{Computation of the mean of the trapped particles}

In order to compute $\frac{1}{2}\int_{-1}^1 f^t d\mu$, we use equation \eqref{f_t_is_beta_t} to simplify the notation. We start by discretizing equation \eqref{eq:trappedR}, implicitly for stability reasons, by
 \begin{equation*}
 \frac{f_i^{t, n+1}- f_i^{t,n}}{c\Delta t} = j_i^{n+1} -\tilde \chi_i^{n+1}f_i^{t,n+1} - \Sigma_i^{n+1}\,,
 \end{equation*}
which, by eliminating $f_i^{t,n+1}$ on the right hand side, can be rewritten as
 \begin{equation}\label{fn1}
  \frac{f_i^{t, n+1}- f_i^{t,n}}{c\Delta t} = \frac{j_i^{n+1}-\tilde \chi_i^{n+1}f_i^{t,n}-\Sigma_i^{n+1}}{1+ \tilde \chi_i^{n+1}c\Delta t}\,.
 \end{equation}
 
 For ease of notation, we now define $\alpha:=  \frac{1}{r^2}\frac{\partial}{\partial r} \left( -r^2\frac{\lambda}{3} \frac{\partial f^t}{\partial r} \right)$ as the diffusion part of the diffusion source $\Sigma$. 
We discretize $\alpha$ by two centered finite differences as
$$
\alpha_i 
=\frac{-1}{3r_i^2}\frac{\left[r_{i+\frac{1}{2}}^2\lambda_{i+\frac{1}{2}}\frac{f_{i+1}^t-f_i^t}{r_{i+1}-r_i}-r_{i-\frac{1}{2}}^2\lambda_{i-\frac{1}{2}}\frac{f_{i}^t-f_{i-1}^t}{r_{i}-r_{i-1}} \right]}{(r_{i+\frac{1}{2}}-r_{i-\frac{1}{2}})}\,.
$$
We write this expression more compactly as
\begin{equation}
\label{alpha}
\alpha_i = -\xi_i (f_{i+1}^t - f_i^t) + \zeta_i (f_i^t- f_{i-1}^t)\,,
\end{equation}
with the two quantities $\xi_i$, $\zeta_i$ defined by
 \begin{equation*}
 \xi_i = \frac{1}{3r_i^2(r_{i+\frac{1}{2}}-r_{i-\frac{1}{2}})}\frac{r_{i+\frac{1}{2}}^2\lambda_{i+\frac{1}{2}}}{r_{i+1}-r_{i}}\,, \quad
  \zeta_i = \frac{1}{3r_i^2(r_{i+\frac{1}{2}}-r_{i-\frac{1}{2}})}\frac{r_{i-\frac{1}{2}}^2\lambda_{i-\frac{1}{2}}}{r_{i}-r_{i-1}}\,.\\
 \end{equation*}
Consequently, the as yet unlimited $\Sigma_i^{n+1}$, written as $\tilde\Sigma_i^{n+1}$, is given by
 \begin{equation}\label{tSigmaD}
 \tilde \Sigma_i^{n+1} = \alpha_i^{n+1}+\tilde\chi_i^{n+1}\left(\frac{1}{2}\int_{-1}^{1} f^s d\mu\right)_i^{n+1}\,,
 \end{equation}
with
  \begin{equation}\label{alphaD}
 \alpha_i^{n+1} = -\xi_i^n ( f_{i+1}^{t,n} - f_i^{t,n}) + \zeta_i^n(f_i^{t,n+1} - f_{i-1}^{t,n+1})\,,\\[1mm]
 \end{equation}
which is a semi-implicit discretization. We first perform all the computations without the limiters and apply the limiting only at the end of the computations. In this spherically symmetric example the diffusive fluxes propagate almost exclusively outwards, hence we choose to discretize explicitly the inward flux and implicitly the outward flux in \eqref{alphaD}. 

 Inserting the equation \eqref{alphaD} into equation \eqref{tSigmaD} gives
 {{{}}
 \begin{equation}\label{tSigma2}
 \tilde \Sigma_i^{n+1} = -\xi_i^n ( f_{i+1}^{t,n} - f_i^{t,n}) + \zeta_i^n(f_i^{t,n+1} - f_{i-1}^{t,n+1}) + \tilde\chi_{i}^{n+1}\left(\frac{1}{2}\int_{-1}^{1} f^s d\mu\right)_i^{n+1}\,.
 \end{equation}}
 We compute the solution cell by cell from $r=0$ to $r=R$. Therefore, we can assume that all the $i-1$ indexed quantities are known in equation~\eqref{tSigma2}. The only term on the right hand side of~\eqref{tSigma2} that is still unknown is~$f_{i}^{t,n+1}$. To eliminate it we introduce the vanishing term $-\zeta_i^n f_i^{t,n}+\zeta_i^n f_i^{t,n}$ in equation~\eqref{tSigma2} that leads to
 \begin{equation*}
  \tilde \Sigma_i^{n+1} = -\xi_i^n ( f_{i+1}^{t,n} - f_i^{t,n}) + \zeta_i^n(f_i^{t,n+1}- f_i^{t,n})+\zeta_i^n ( f_i^{t,n} - f_{i-1}^{t,n+1}) + \tilde\chi_{i}^{n+1}\left(\frac{1}{2}\int_{-1}^{1} f^s d\mu\right)_i^{n+1}\,,
 \end{equation*}
 and eliminate the second term using equation~\eqref{fn1} with the unlimited diffusion source $\tilde\Sigma_i^{n+1}$. We obtain
  \begin{equation*}
 \tilde \Sigma_i^{n+1} =  \zeta_i^nc\Delta t\left(\frac{j_i^{n+1}-\tilde \chi_i^{n+1}f_i^{t,n}-\tilde\Sigma_i^{n+1}}{1+ \tilde \chi_i^{n+1}c\Delta t}\right) -\xi_i^n ( f_{i+1}^{t,n} - f_i^{t,n}) +\zeta_i^n ( f_i^{t,n} - f_{i-1}^{t,n+1}) + \tilde\chi_{i}^{n+1}\left(\frac{1}{2}\int_{-1}^{1} f^s d\mu\right)_i^{n+1}\,.
 \end{equation*}
Solving this equation for $\tilde \Sigma_i^{n+1}$ gives
 \begin{equation*}
 \begin{aligned}
 \tilde \Sigma_i^{n+1} = &\ \frac{1}{1 + (\zeta_i^n + \tilde\chi_i^{n+1})c\Delta t}\bigg\{ \zeta_i^n c\Delta t(j_i^{n+1} - \tilde \chi_i^{n+1}f_{i}^{t,n})  \\
 &+ (1+ \tilde \chi_i^{n+1}c\Delta t) \cdot \bigg[ -\xi_i^n ( f_{i+1}^{t,n} - f_i^{t,n}) +\zeta_i^n ( f_i^{t,n} - f_{i-1}^{t,n+1})\\
 &+ \tilde \chi_i^{n+1}\left(\frac{1}{2}\int_{-1}^{1} f^s d\mu\right)_i^{n+1} \bigg] \bigg\}\,.
 \end{aligned}
 \end{equation*}
We can now apply the limiter
 \begin{equation*}
 \Sigma_i^{n+1} = \min \left\{ \max\left[ \tilde \Sigma_i^{n+1},\ 0 \right],\  j_i^{n+1}\right\}\,.\\[2mm]
 \end{equation*}
 The updated diffusion source $\Sigma_i^{n+1}$ can now be used in the computation of the updated $f_i^{t,n+1}$ using equation~\eqref{fn1}.

\section{Numerical results}
\label{numericalresSec}

In this section, we study the discrete reduced model and the discretized IDSA numerically. First, we perform a numerical convergence study in order to validate
the discretizations and obtain concrete truncation error estimates for
the study of the modeling error. Next, we illustrate the different
behaviours of the two discretizations for a concrete example, for which
we also quantify the modeling error of the IDSA.

\subsection{Study of the discretization error}\label{DiscrErr}
We study the discretization error in two different ways for the two models. In the case of the reduced Boltzmann problem, we pick a solution, define the corresponding $j$ and $\tilde\chi$ and compare the computed solution with the exact one. In the IDSA case, however, this strategy does not work because we cannot find a simple solution of the coupled system. Therefore, we compare the solutions on different refinement levels with a reference solution on the most refined grid. Concretely, we choose $10$ refinements of the coarse grid for the reference solution and, in order to compute the discretization order, compare the solutions obtained by $0,\ldots,8$ refinements with it. The parameters used in this study correspond to those used in the numerical example shown later in Subsection~\ref{num}.

To study the discretization error on the domain $(r,\mu)\in \Omega = [0,10]\times[-1,1]$, we start from a coarse grid with $N_r=5$ discretization points in the $r$-direction, $N_{\mu}=3$ discretization points in the $\mu$-direction and with $N_{t}=1$ time steps. We choose a time interval $[0,T]$ where $T = 0.0005$ for the Boltzmann simulation and $T=0.1$ for the IDSA one. 

Since we expect both discretizations to be of order one both in time and space, at each step we refine the grid by dividing the step sizes $\Delta t$, $\Delta r$ and $\Delta\mu$ by two. Therefore, these step sizes are proportional. The error is computed with the solutions obtained at the last time step.

Concretely, in order to study the discretization error of the reduced Boltzmann model, we choose the analytical solution
\begin{equation*}
f_{\mbox{\tiny exact}} := \frac{(1-t)r}{1050}\left[(r-10)^2+(\mu+1)^2+1 \right]\, ,\\[2mm]
\end{equation*}
which represents a paraboloid in the domain $(t,r,\mu)\in[0,1]\times\Omega$. The normalization constant $1050$ has been chosen to insure $f_{\mbox{\tiny exact}}(r,\mu,t)<1$. 

With the choice of the values for the emissivity $j_{\mbox{\tiny exact}}$ and the opacity $\tilde\chi_{\mbox{\tiny exact}}$ as
\begin{equation*}
j_{\mbox{\tiny exact}} :=\frac{(1-t)}{1050}\left\{\mu\left[(r-10)^2+(\mu+1)^2+1\right]+2(1-\mu^2)(\mu+1)\right\}\, ,
\end{equation*}
and
\begin{equation*}
\tilde\chi_{\mbox{\tiny exact}} := \frac{\left[(r-10)^2+(\mu+1)^2+1\right]-2\mu(1-t)(r-10)}{(1-t)[(r-10)^2+(\mu+1)^2+1]}\,, \\[1mm]
\end{equation*}
the function $f_{\mbox{\tiny exact}}$ is a solution of equation \eqref{conserv}.

In this setting we now let $T = 0.0005$. By Remark~\ref{FlowCFL} we need a very small value of $\Delta t$ because we have $\Delta t = T$ in the first iteration. It would be too costly to compute the order for $T=0.1$ for this example because of the restrictive CFL condition.

\begin{figure}[h!]
\centering
\subfigure[Reduced Boltzmann model with $T = 0.0005$: The dependency of the error in the infinity norm on $\Omega$ on the spatial mesh size $h=\Delta r$ is displayed. The curve with round markers represents the relative error, the curve with the plus markers displays the absolute error. The straight line has slope one.]{\includegraphics[width = .4\linewidth]{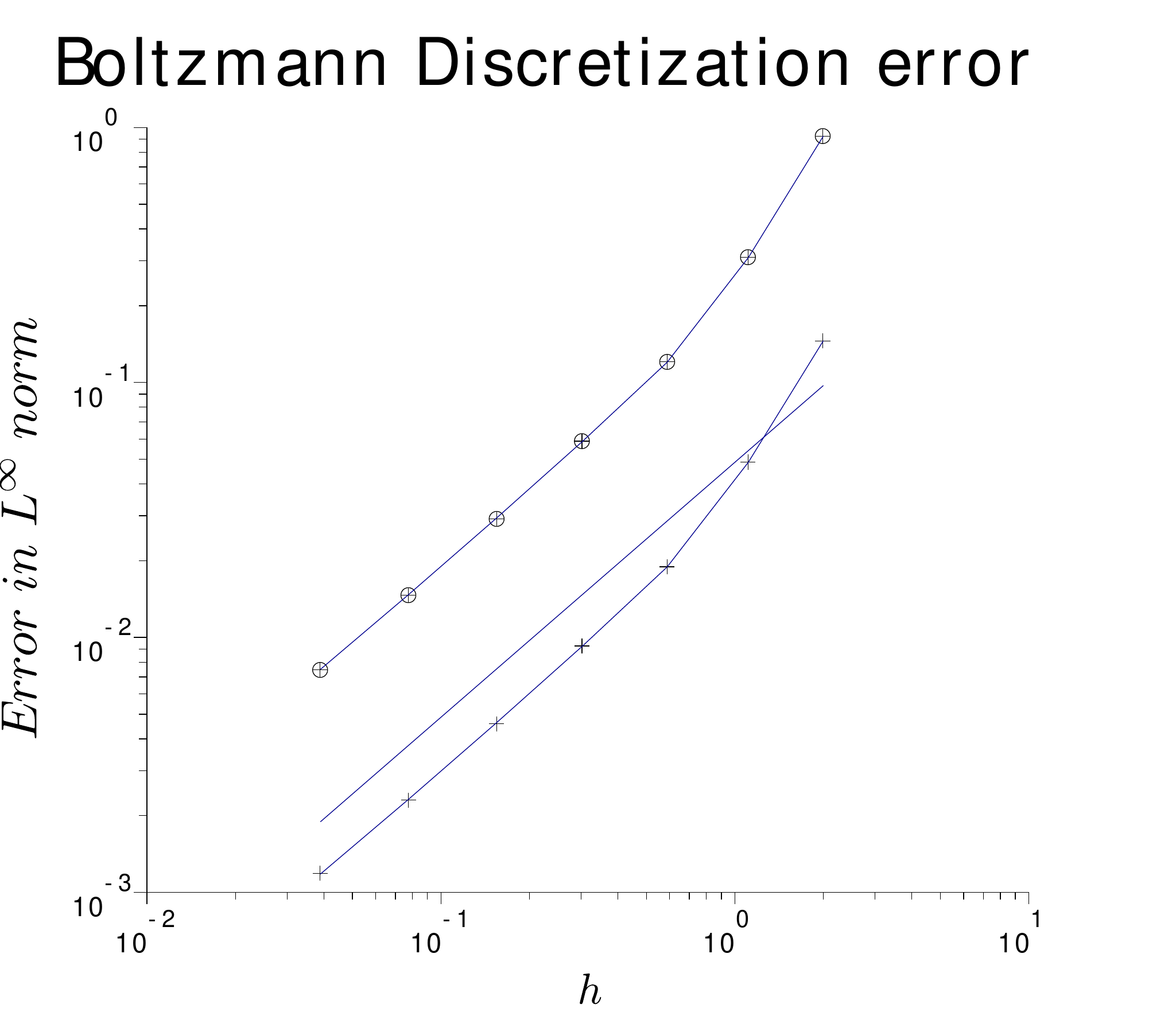}}\qquad
\subfigure[IDSA model with $T = 0.1$: The dependency of the error in the infinity norm on {$[0,10]$} on the spatial mesh size $h=\Delta r$ is displayed. In this case the absolute and the relative error are equal because the infinity norm of $f$ is $1$. The straight line has slope one.]{\includegraphics[width = .4\linewidth]{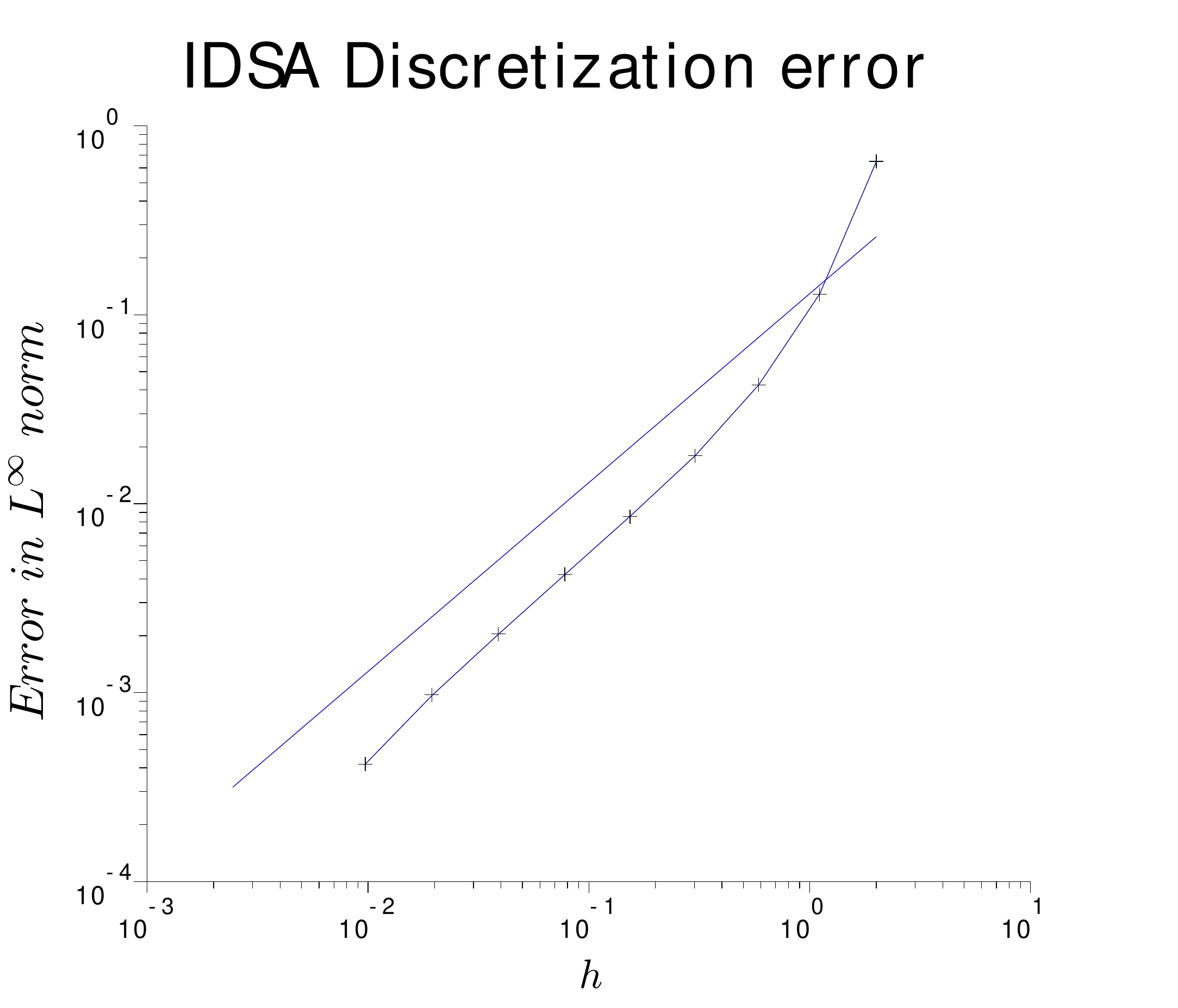}}
\caption{Discretization error results for the reduced Boltzmann model and IDSA.}
\label{DiscrError}
\end{figure}

Figure \ref{DiscrError} shows the discretization error in the infinity norm on $\Omega=[0,10]\times [-1,1]$ for the reduced Boltzmann model in panel (a) and on $[0,10]$ for the IDSA model in panel (b). In both cases, the error is of order one as one would expect for simple linear cases with the discretizations used.
\vspace{-0.2cm}
\subsection{Comparison of the reduced model and the IDSA}

In this section, we present a comparison of the reduced Boltzmann model and the IDSA by a numerical example and provide a study of the modeling error with the help of this example. 

\subsubsection{Numerical example}\label{num}
The numerical example we study is characterized by the following parameters. We use a domain $(r,\mu)\in\Omega_{\mbox{\tiny Bol}} = [0,R]\times [-1,1]$ with $R=9$ and $T=20$. The domain of computation $r \in\Omega_{\mbox{\tiny IDSA}} = [0,R]$ for the IDSA is different because we do not have the dependency on $\mu$. The grid used here has $N_r =100$ discretization points in the $r$-direction and $N_\mu = 20$ discretization points in the $\mu$-direction. The time step is $\Delta t = 0.1$. This time is small enough with regard to the CFL condition, because we force the region where the CFL condition is critical to be dominated by reactions, see Remark \ref{FlowCFL}.

We choose the space interval to be $[0,9]$ and divide the computational domain into three parts in order to see the behaviour of the different (reaction, diffusion, free streaming) regimes of the Boltzmann equation that are representable in the IDSA. In our example, the occurrence of the different regimes depends on the values of the emissivity $j(r)$ and of the opacity $\tilde\chi(r)$ that are linked by a Fermi-Dirac distribution. We define the emissivity to be $j(r)=1$ on $[0,3]$, $j(r)=10^{-3}$ on $[6,9]$, linear on $[3,6]$ and continuous on $[0,9]$. This choice leads to reaction dominance on $[0,3]$ and dominance of transport on $[6,9]$. In the rest of the computational domain we have the transition between the two regimes. We fix the Fermi-Dirac distribution to be $(\exp(-20)+1)^{-1}$, which implies $\tilde\chi=j\,(\exp(-20)+1)$ for the opacity. Finally, we set the neutrino scattering sphere radius $R_{\nu} = 4.5$.

We use homogeneous Neumann boundary conditions at $r = 0$ if $\mu\leq 0$ and at $r=9$ if $\mu\geq 0$ and homogeneous Dirichlet boundary conditions at $r = 0$ if $\mu> 0$ and at $r=9$ if $\mu< 0$. These boundary conditions are set to match the flow of transport. Dirichlet conditions correspond to incoming flow and Neumann conditions correspond to outgoing flow, see Figure~\ref{flowGraph}. The Neumann conditions are discretized by creating ghost cells just outside the domain that take the same value as the previous ones. For the IDSA we use the fact that we impose different regimes in different parts of the domain. We therefore use homogeneous Neumann boundary conditions at $r=0$ for $f^t$ and at $r =9$ for $f^s$ and Dirichlet boundary conditions for the other two : $f^s(0) = 0$ and $f^t(9) = 0$. 
\begin{rmrk} We do not need any boundary conditions in the $\mu$-direction because the flux vanishes along these boundaries, see Subsection~\ref{flow}.
\end{rmrk}

As initial condition we use $f(r,\mu,0) = \frac{1}{2}$ for the reduced Boltzmann model. Motivated by the choice of $j(r)$ we choose $f^t$ to be $f^t(r,0) = \frac{1}{2}$ on $[0,3]$, $f^t(0)=0$ on $[6,9]$, linear on $[3,6]$ and continuous on $[0,9]$ as well as $f^s(r,0) = \frac{1}{2}-f^t(r,0)$.

\begin{rmrk}\label{unPhysCond} A constant initial condition is not physical. In a spherical domain, we expect a decrease of the distribution function $f$ with respect to the radius $r$. However, this is not a big issue since the reaction part will force any solution to converge to the correct one and the transport part will eventually remove the rest of the unphysical initial condition.
\end{rmrk}

\begin{figure}[h!]
\centering
\subfigure[Boltzmann distribution function in the phase space {$(r,\mu)\in [0,9]\times [-1,1]$}.]{
\includegraphics[width = .45\linewidth]{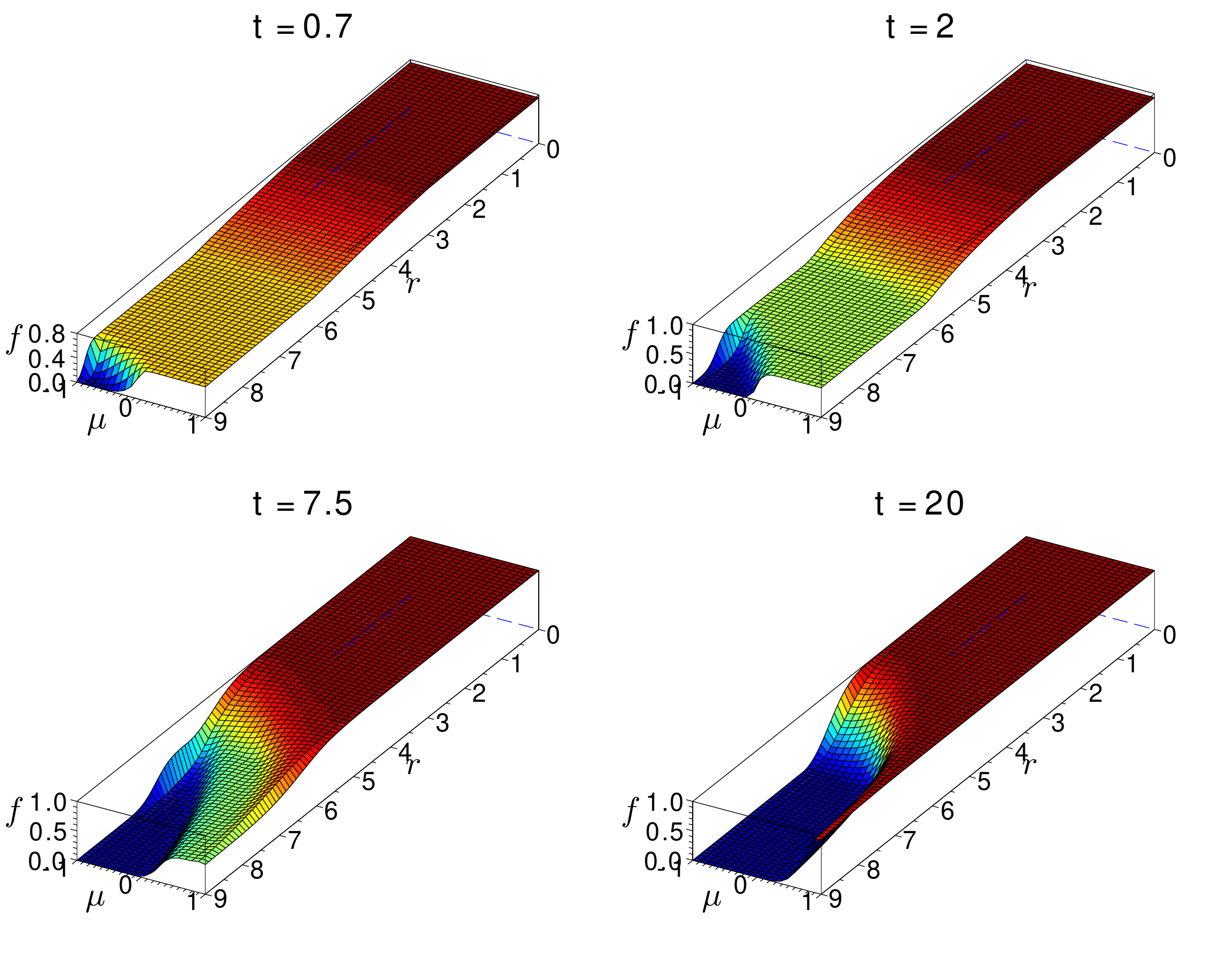}
}\quad
\subfigure[Comparison of the averaged distribution functions. The solid blue line represents the reduced Boltzmann solution. The blue dashed line represents the IDSA solution, which is the sum of the red dashed line representing the trapped component and the green dashed line representing the free streaming component.]{
\includegraphics[width = .45\linewidth]{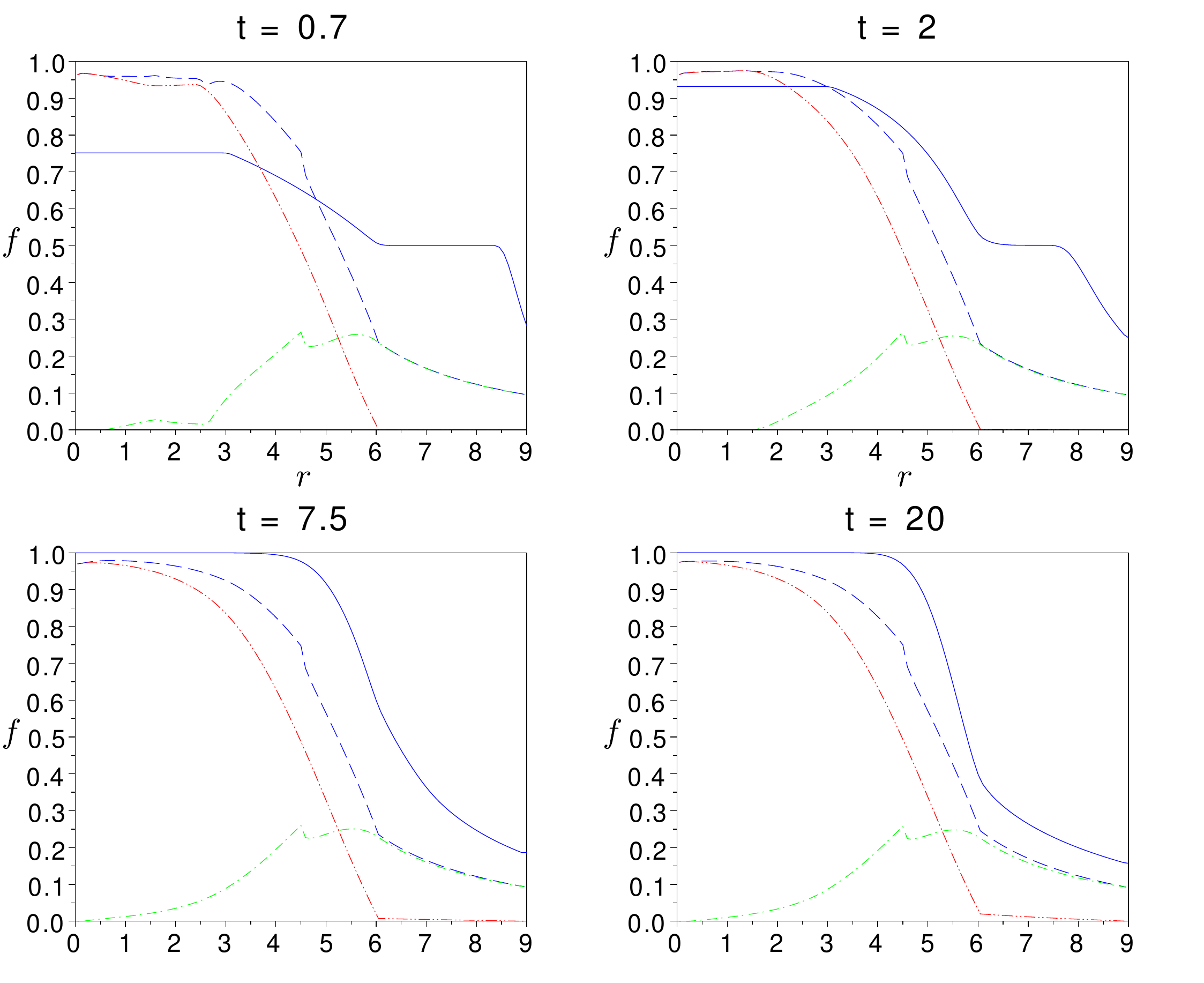}
}
\caption{This figure shows 4 snapshots of the evolution of the Boltzmann distribution function $f$ in the phase space (panel (a)) and the same for its angular average compared to the IDSA solution (panel (b)) at times t = 0.7, 2, 7.5 and 20.}
\label{Short}
\end{figure}

The results of the simulation are shown in Figure \ref{Short} which displays four relevant snapshots of the evolution of the distribution function (panel (a)) and of the angular mean of it compared to the IDSA solution (panel (b)).
Panel (a) of Figure \ref{Short} shows the distribution function at four chosen times. In the subdomain $[0,3]\times[-1,1]$, the evolution is driven by the reactions and, therefore, the only effect that we see is an isotropic growth of the function towards the equilibrium $f_{eq} = j/\tilde\chi = (\exp(-20)+1)^{-1} \approx 1$. In the subdomain $[6,9]\times[-1,1]$, the evolution is dominated by transport. It is therefore not isotropic and the neutrinos are moved along their trajectory lines as shown in Figure \ref{flowGraph}. As explained in Remark \ref{unPhysCond}, the unphysical initial condition is eventually removed. As the flow of transport does not have any effect on an isotropic zone because it is conservative, we start to see its effect on the boundary at $r=9$ and for $\mu<0$. The fact that $\mu$ is negative reflects incoming neutrinos. As we set the distribution function to be zero outside the domain, the isotropy reduces the distribution function. The transport propagates in the subdomain as time evolves and removes the unphysical initial condition after a finite time. Another anisotropy is propagating in this domain, arising from the center reaction dominated zone. This zone acts as a source of neutrinos that propagate outwards, into the $\mu>0$ subdomain. We start to see this effect in the $t=2$ snapshot, and it continues to grow in the other two. At the end of this simulation, the unphysical initial condition has been completely removed, the reaction dominated region is equilibrated and produces neutrinos that propagate mainly outwards, driven by the flow of transport. The intermediate region just shows how the reaction dominated region is coupled with the transport dominated region.

Panel (b) of Figure \ref{Short} shows the comparison between the two models at the same four times as panel (a). In order to discuss the comparison, we divide the domain into the 3 subdomains as before, concretely into $\Omega_{\mbox{\tiny reac}}= [0,3]$, the subdomain dominated by reactions, $\Omega_{\mbox{\tiny transp}} =[6,9]$, the subdomain dominated by transport and the intermediate subdomain $\Omega_{\mbox{\tiny int}} = (3,6)$.

In $\Omega_{\mbox{\tiny reac}}$ we see that the convergence of the IDSA solution $f_{\mbox{\tiny IDSA}}$ to the equilibrated one is much faster than the convergence of solution $f_{\mbox{\tiny Bol}}$ for the Boltzmann model. This is explained by the fact that the trapped component of $f_{\mbox{\tiny IDSA}}$ is strongly coupled with the free streaming component which evolves infinitely fast since we use a stationary state approximation. The coupling between the two components of $f_{\mbox{\tiny IDSA}}$ therefore explains the faster convergence rate of it to equilibrium. As seen in the discussion of panel (a), in this domain, the distribution function $f_{\mbox{\tiny Bol}}$ is isotropic and, therefore, its angular mean evolves in the same way. We notice that the IDSA is underestimating the Boltzmann solution. 

In $\Omega_{\mbox{\tiny transp}}$ we see that the angular mean of $f_{\mbox{\tiny Bol}}$ describes the reduction of the unphysical initial condition as explained in the discussion of panel (a). In the two first graphs at times $t=0.7$ and $t=2$, we see some regions that are still representing the isotropic initial condition. In the two other graphs, we do not notice its presence anymore even if there is still a component of it at $t=7.5$ as shown in panel (a). For the IDSA we see, as expected, that this region is described by the streaming component of $f_{\mbox{\tiny IDSA}}$, but it shows an underestimation of the reference Boltzmann solution. We think that this should be explained by the way of coupling the two regions $\Omega_{\mbox{\tiny reac}}$ and $\Omega_{\mbox{\tiny transp}}$.

In the intermediate subdomain $\Omega_{\mbox{\tiny int}}$, we see the evolution of the transition between the two other subdomains, in particular, the transition between the trapped and the streaming components of $f_{\mbox{\tiny IDSA}}$. As expected, we see the growth of the streaming part in the intermediate domain and a decrease in the transport domain. There are two aspects that we want to point out here. First, we see a strange behaviour of the streaming component around $r=4.5$. This behaviour is a consequence of the neutrino scattering sphere radius that has been set precisely to this value of $r$. Second, we observe that the streaming component does not vanish in the reaction dominated regime as expected. There is a noticeable component until $r\approx 1$ and, as a consequence of the coupling, this reduces a little the trapped component which, then, underestimates the equilibrium value.

\subsubsection{Study of the modeling error in IDSA}

In this subsection and on the basis of the numerical example of the previous subsection, we study the modeling error of the isotropic diffusion source approximation compared to the Boltzmann model. In order to have a reliable measure of this error, we need to control the discretization error. Using the results of Section~\ref{DiscrErr}, we choose the grid parameters such that the relative discretization error is less than one percent. The parameters are $N_r = 257$, $N_\mu=129$, and $\Delta t = 1.5625\cdot 10^{-3}$.
\begin{figure}[h!]
\centering
\subfigure[IDSA modeling error, computed as the relative error in the $L^2$ norm of $f_{\mbox{\tiny IDSA}}$ to $f_{\mbox{\tiny Bol}}$.]{
\includegraphics[width = .47\linewidth]{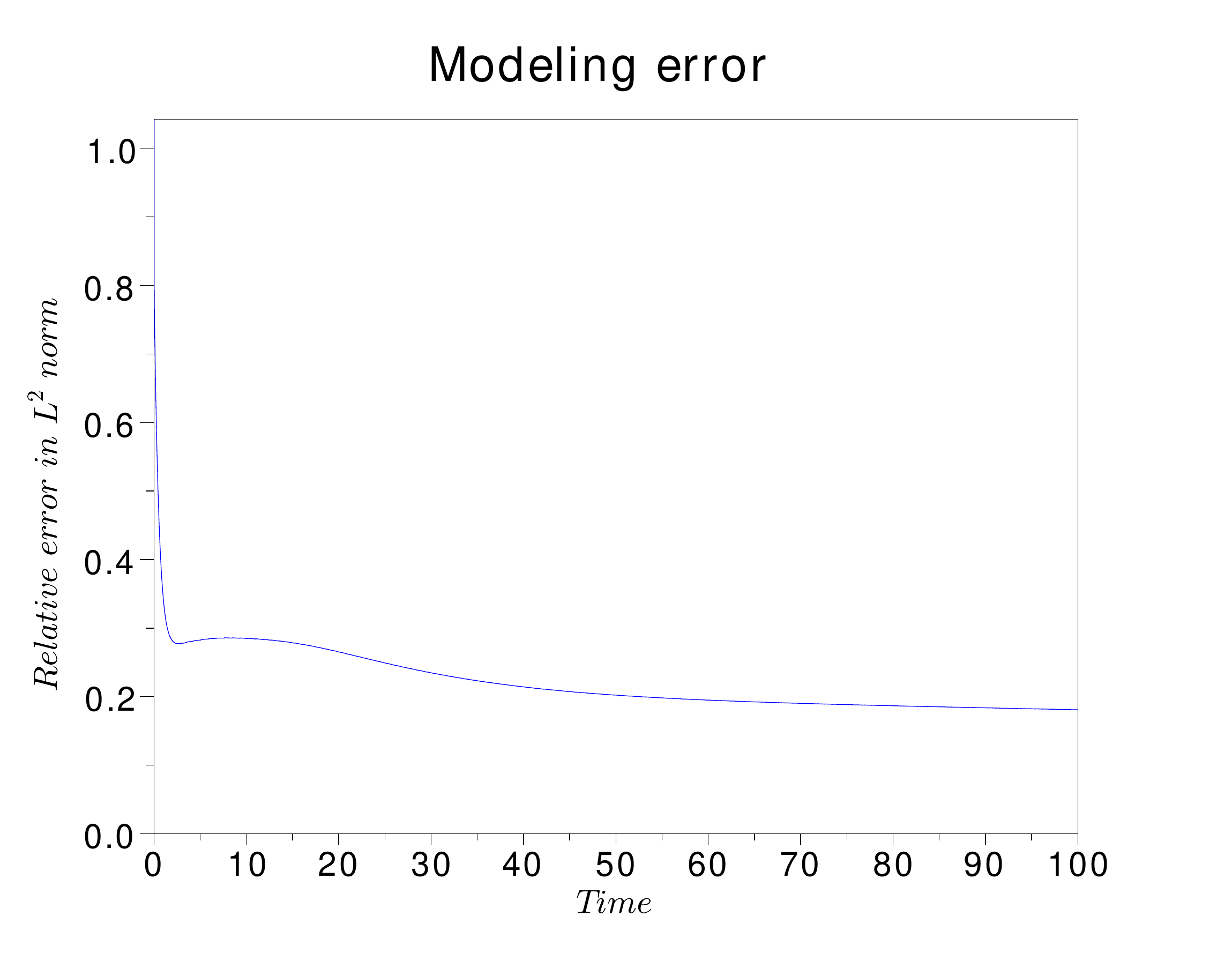}
}\quad
\subfigure[Localization of the error at time t=100, computed by the absolute difference of the two solutions $|f_{\mbox{\tiny Bol}}-f_{\mbox{\tiny IDSA}}|$.]{
\includegraphics[width = .47\linewidth]{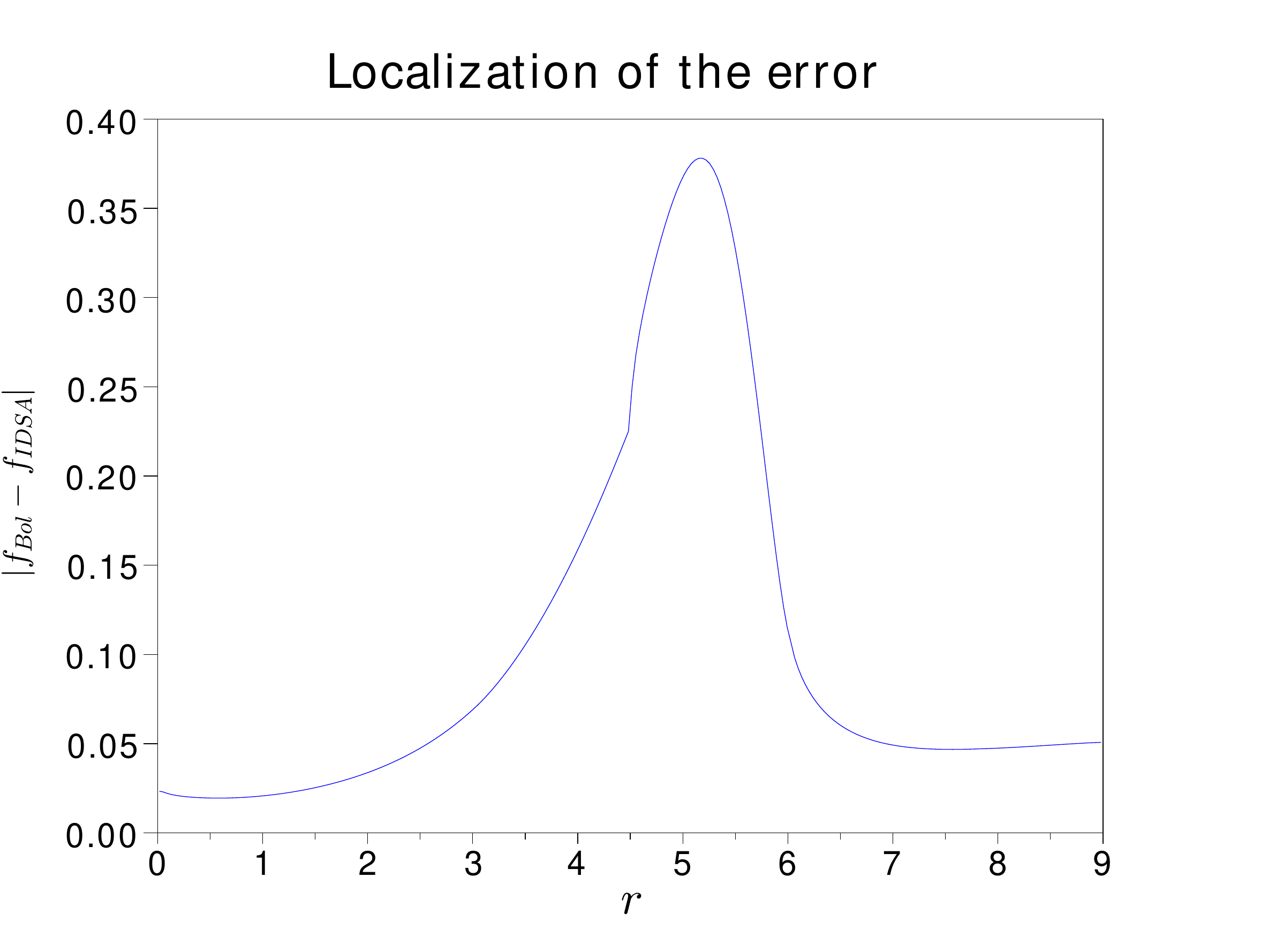}
}
\caption{Modeling error of the IDSA.}
\label{modelingError}
\end{figure}

The results shown in Figure \ref{modelingError} quantify the error of the IDSA with respect to the Boltzmann model. As we fix the discretization error to be smaller than one percent, we know that it contributes less than $0.02$ in the graph of panel (a). We can therefore conclude that the relative modeling error is $20\%\pm 2\%$ which one might find quite big. A relevant question to ask is if the error is uniform or accumulates in some regions. To answer this question, we compute in each spatial point the absolute difference between the two solutions. The result is shown in panel (b) of Figure \ref{modelingError}. As expected, at the end of the simulation, the error is mainly located in the region of transition $(3,6)$ between the reaction and free streaming regimes. 

Summarizing, this preliminary study shows that the IDSA is qualitatively reasonable, but it also shows that the coupling between the two regimes exhibits considerable errors and is the main source of error of the IDSA. 

\bibliographystyle{plain}
\bibliography{Proceeding}

\end{document}